\documentclass[useAMS,usenatbib,twocolumn]{mnras}
\usepackage{graphicx}
\usepackage{amsmath}
\usepackage{amssymb}
\usepackage{xspace}
\usepackage{commath}
\usepackage{times}
\usepackage{bm} 
\usepackage{balance}
\usepackage{hyperref}
\usepackage{mathtools}
\usepackage{stmaryrd}
\usepackage{tabulary}

\hypersetup{dvips, colorlinks=true, linkcolor=black, citecolor=black, filecolor=blue, urlcolor=black}

\newcommand{\rt}{\mathrm{t}}
\newcommand{\rr}{\mathrm{r}}
\newcommand{\rc}{\mathrm{c}}
\newcommand{\ra}{\mathrm{a}}
\newcommand{\rp}{\mathrm{p}}
\newcommand{\rb}{\mathrm{b}}
\newcommand{\rd}{{\mathrm{d}}}
\newcommand{\rD}{{\mathrm{D}}}
\newcommand{\rHU}{{\mathrm{HU}}}
\newcommand{\Rv}{R_{\mathrm{v}}}
\newcommand{\sgn}{\mathrm{sgn}}
\newcommand{\tmax}{t_{\mathrm{max}}}
\newcommand{\tlast}{t_{\mathrm{last}}}
\newcommand{\Nrun}{N_{\mathrm{run}}}
\newcommand{\Nbound}{N_{\mathrm{bound}}}
\newcommand{\Navg}{N_{\mathrm{avg}}}
\newcommand{\kmin}{k_{\mathrm{min}}}
\newcommand{\kmax}{k_{\mathrm{max}}}
\newcommand{\Lmax}{L_{\mathrm{max}}}
\newcommand{\Rmax}{R_{\mathrm{max}}}

\newcommand{\sint}{{\!\int\!}}
\newcommand{\p}{\partial}

\newcommand{\Phieff}{\Phi_{\mathrm{eff}}}
\newcommand{\rra}{r_{\ra}}
\newcommand{\rrp}{r_{\rp}}
\newcommand{\vr}{v_{\rr}}
\newcommand{\vt}{v_{\rt}}

\newcommand{\vx}{v_{x}}
\newcommand{\vy}{v_{y}}
\newcommand{\vz}{v_{z}}

\newcommand{\vpp}{v^{\prime}}
\newcommand{\vppr}{\vr^{\prime}}
\newcommand{\vppt}{\vt^{\prime}}
\newcommand{\vppx}{v^{\prime}_{x}}
\newcommand{\vppy}{v^{\prime}_{y}}
\newcommand{\vppz}{v^{\prime}_{z}}

\newcommand{\vppSq}{v^{\prime 2} }
\newcommand{\vpprSq}{\vr^{\prime 2}}
\newcommand{\vpptSq}{\vt^{\prime 2}}

\newcommand{\mb}{m_{\rb}}
\newcommand{\Eo}{E_{0}}
\newcommand{\Epp}{E'}
\newcommand{\Lo}{L_{0}}
\newcommand{\Omegao}{\Omega_{0}}
\newcommand{\Lpp}{L'}
\newcommand{\Jr}{J_{\rr}}
\newcommand{\sra}{s_{\ra}}
\newcommand{\srp}{s_{\rp}}
\newcommand{\sst}{s_{\star}}
\newcommand{\Rc}{R_{\rc}}
\newcommand{\deltaD}{\delta_{\rD}}
\newcommand{\res}{\mathrm{res}}

\newcommand{\tE}{\widetilde{E}}
\newcommand{\tL}{\widetilde{L}}
\newcommand{\hu}{\widehat{u}}

\newcommand{\Ftot}{F_{\mathrm{tot}}}
\newcommand{\mH}{\mathbb{H}}
\newcommand{\HG}{\prescript{}{2}F_{1}}

\newcommand{\e}[1]{\boldsymbol{{\mathrm{e}}}_{#1}}
\newcommand{\bvpp}{\boldsymbol{v}^{\prime}}
\newcommand{\bv}{\boldsymbol{v}}
\newcommand{\hv}{\hat{\bv}}
\newcommand{\bvt}{\boldsymbol{v}_{\mathrm{t}}}
\newcommand{\bw}{\boldsymbol{w}}
\newcommand{\bk}{\boldsymbol{k}}
\newcommand{\bF}{\boldsymbol{F}}
\newcommand{\bJ}{\boldsymbol{J}}
\newcommand{\br}{\boldsymbol{r}}
\newcommand{\hr}{\hat{\br}}
\newcommand{\bD}{\boldsymbol{D}}
\newcommand{\brc}{\boldsymbol{r}_{\mathrm{c}}}

\newcommand{\dvPar}{\Delta v_{\parallel}}
\newcommand{\dvPerp}{\Delta v_{\perp}}
\newcommand{\dvParAvrLoc}{\langle \dvPar \rangle}
\newcommand{\dvParSqAvrLoc}{\langle(\dvPar)^{2}\rangle}
\newcommand{\dvPerpSqAvrLoc}{\langle(\dvPerp)^{2}\rangle}
\newcommand{\DE}{\Delta E}
\newcommand{\DL}{\Delta L}
\newcommand{\DEAvr}{\langle\DE\rangle}
\newcommand{\DESqAvr}{\langle (\DE)^{2} \rangle}
\newcommand{\DLAvr}{\langle\DL\rangle}
\newcommand{\DLSqAvr}{\langle (\DL)^{2} \rangle}
\newcommand{\DEDLAvr}{\langle \DE \DL \rangle}

\newcommand{\half}{\tfrac{1}{2}}

\usepackage{soul} \usepackage{xcolor}
\definecolor{aquamarine}{rgb}{0.5, 1.0, 0.83}

\begin{document}

\title[Non-resonant relaxation]{Non-resonant relaxation of anisotropic globular clusters}

\author[K. Tep, J.-B. Fouvry, C. Pichon]{Kerwann Tep$^{1}$, Jean-Baptiste Fouvry$^{1}$ and Christophe Pichon$^{1,2}$\\
\noindent
$^{1}$ CNRS and Sorbonne Universit\'e, UMR 7095, Institut d'Astrophysique de Paris, 98 bis Boulevard Arago, F-75014 Paris, France\\
\noindent$^{2}$ IPhT, DRF-INP, UMR 3680, CEA, Orme des Merisiers B\^at 774, F-91191 Gif-sur-Yvette, France
}

\maketitle

\begin{abstract}
Globular clusters are dense stellar systems
whose core slowly contracts under the effect of self-gravity.
The rate of this process was recently found to be directly linked to the initial amount of velocity anisotropy: tangentially anisotropic clusters contract faster than radially anisotropic ones. Furthermore, initially anisotropic clusters are found
to generically tend towards more isotropic distributions during the onset of contraction.
Chandrasekhar's ``non-resonant'' (NR) theory of diffusion
describes this relaxation as being driven by
a sequence of local two-body deflections along each star's orbit.
We explicitly tailor this NR prediction to anisotropic clusters,
and compare it with $N$-body realisations
of Plummer spheres with varying degrees of anisotropy.
The NR theory is shown to recover remarkably well
the detailed shape of the orbital diffusion
and the associated initial isotropisation,
up to a global multiplicative prefactor
which increases with anisotropy.
Strikingly, a simple effective isotropic prescription
provides almost as good a fit,
as long as the cluster's anisotropy is not too strong.
For these more extreme clusters,
accounting for long-range resonant relaxation
may be necessary to capture these clusters' long-term evolution.
\end{abstract}


\section{Introduction}
\label{sec:intro}

Understanding the long-term evolution of globular clusters is a long-standing problem in stellar dynamics~\citep{Henon1961,Harris+1979,Spitzer1987}.
Not only is the dynamics of globular clusters
interesting \textit{per se}~\citep[see, e.g.\@,][]{Lightman+1978,Harris1991,Meylan+1997,Brodie+2006},
but it is also the archetype for the relaxation
of a (weakly) collisional self-gravitating system
with a simple integrable configuration~\citep[see, e.g.\@,][for a review]{Chavanis2013a}.

In~\cite{Chandrasekhar1943}'s picture,
the velocity of a given test star undergoes a series of weak, local, and uncorrelated kicks from each field star it encounters,
a process that we coin ``non-resonant relaxation'' (NR).
Once these deflections accumulated along the stars'
underlying unperturbed orbits,
NR provides us with the classical picture for long-term relaxation in spherical clusters~\citep{HeggieHut2003}.
In practice, NR is rather straightforward to implement
for isotropic clusters and has been extensively used to describe their long-term evolution~\citep[see, e.g.\@,][for a review]{Vasiliev2015}.
The same approach was also recently updated via the (inhomogeneous) Balescu--Lenard equation~\citep{Heyvaerts2010,Chavanis2012} 
to account for gravitational wakes and large-scale resonances
within the globular clusters~\citep{Hamilton+2018,Fouvry+2021}.
Overall these non-local effects were, somewhat surprisingly,
found to be of small relevance for such isotropic spheres.
Moving away from isotropy, one could expect
that more coherent motions within the cluster,
e.g.\@, via velocity anisotropies,
could affect these systems' long-term evolution.
In this paper, we wish to quantify the extent
to which NR still applies for such systems.

\cite{Longaretti1997,Kim2008,Hong2013}
studied the impact of rotation on globular clusters
using $N$-body simulations and Fokker--Planck models.
They concluded that core collapse was accelerated in clusters
with a non-zero total angular momentum.
For non rotating, but anisotropic clusters, \citet{Cohn1979} devised an orbit-averaged Fokker--Planck equation to integrate self-consistently
the evolution of a spherical star cluster.
In that paper, the velocity diffusion coefficients are computed using a pseudo-isotropic distribution fonction, i.e. at a fixed radius,
the test star is scattered by a locally isotropic background of perturbers.
This approach was further refined in the Fokker--Planck simulations of~\citet{Drukier+1999} to carefully treat the effects of velocity-space anisotropy. More recently, \citet{Breen+2017} used direct $N$-body simulations to investigate the relaxation of isolated equal-mass star clusters, primarily focusing on the effects of primordial velocity anisotropies. Interestingly, collapse is found to be swifter as the model becomes more and more tangentially anisotropic.

The purpose of our paper is to extend the NR theory to anisotropic clusters and to model the result of~\cite{Breen+2017}.
The comparison to tailored simulations also allows us to assess the relative performance of the pseudo-isotropic computation presented in~\cite{Cohn1979}.
The paper is organised as follows.
In~\S\ref{sec:NR} we tailor Chandrasekhar's NR theory to anisotropic spherical clusters,
then in~\S\ref{sec:Application} we apply this approach to Plummer spheres, 
while we discuss our results and conclude in~\S\ref{sec:discussion}.

\section{Non-resonant relaxation}
\label{sec:NR}

We consider a self-gravitating globular cluster composed of $N$
stars of individual mass $\mb = M / N$,
with $M$ the cluster's total mass.
Assuming that this cluster is in a quasi-stationary equilibrium,
we characterise its phase-space statistics using the total DF,
$\Ftot = \Ftot (\br , \bv)$, with $\br$ the position and $\bv$
the velocity, normalised so that $\sint \rd \br \rd \bv \Ftot = M$.

\subsection{Local velocity diffusion coefficients}
\label{subsec:loc_vel_diff_coeffs}

As a result of the cluster's finite number of constituents,
a given test star of mass $m$ and velocity $\bv$,
embedded in such a noisy environment
will irreversibly see its velocity diffuse.
This long-term relaxation is driven by pairwise encounters,
a process that we call non-resonant (NR) relaxation.
More precisely, assuming that the deflection is local and following~\S{7.4.4} of~\cite{Binney2008}
(see also~\cite{Chavanis2013} for a review),
the test star's velocity will locally diffuse according
to the first- and second-order velocity diffusion coefficients
\begin{subequations}
\label{eq:generic_D}
\begin{align}
\langle \Delta v_{i} \rangle &= 4\pi G^2 (m+\mb) \ln \Lambda \, \frac{\partial h}{\partial v_{i}} ,
\\
\langle \Delta v_{i} \Delta v_{j} \rangle &= 4\pi G^2 \mb \, \ln \Lambda \, \frac{\partial^2 g}{\partial v_{i}\partial v_{j}} ,
\end{align}
\end{subequations}
where $i$ and $j$ run over the three directions
of the coordinate system.
In that expression, $G$ is the gravitational constant
and $\ln \Lambda$ stands for the Coulomb logarithm
stemming from the heuristic regularisation of local and far-away encounters.
Finally, equation~\eqref{eq:generic_D} involves
the Rosenbluth potentials~\citep{Rosenbluth+1957},
which read
\begin{subequations}
\label{eq:def_gh}
\begin{align}
h(\br , \bv) & = \sint \rd \bvpp \frac{\Ftot(\br,\bvpp)}{ |\bv-\bvpp|} ,
\\
g(\br ,\bv) & = \sint \rd \bvpp\, \Ftot(\br,\bvpp) \, |\bv-\bvpp| .
\end{align}
\end{subequations}
While fully generic, equations~\eqref{eq:generic_D} and~\eqref{eq:def_gh}
are typically further simplified by assuming spherical symmetry and a locally isotropic velocity
distribution~\citep[see, e.g.\@,][]{henon1958}.
In that limit, one imposes $\Ftot (\br , \bv) = \Ftot (r , v)$, with $v=|\bv|$, $r=|\br|$,
and all the integrals from equation~\eqref{eq:def_gh}
become one-dimensional.
One key goal of our paper
is to assess the validity of this isotropy assumption.

\subsection{Anisotropic diffusion coefficients}
\label{subsec:aniso_diff_coeffs}

For a non-rotating anisotropic DF with spherical symmetry, one generically has
$\Ftot (\br , \bv) = \Ftot (r , \vr , \vt)$,
where $\vr$ and $\vt$ are respectively
the radial and tangential velocities,
satisfying $v^2 = \vr^2 + \vt^2$.

In~\S\ref{sec:LocalDiff}, we show that within these coordinates
the diffusion coefficients from equation~\eqref{eq:generic_D}
are fully captured by
\begin{subequations}
\label{eq:generic_grad}
\begin{align}
\dvParAvrLoc =&\,4\pi G^{2}(m+\mb)\ln\Lambda \bigg( \frac{\vr}{v}\frac{\partial h}{\partial \vr}+\frac{\vt}{v}\frac{\partial h}{\partial \vt} \bigg) ,
\\
\dvParSqAvrLoc  =&\, 4\pi G^{2}{\mb}\ln\Lambda\bigg[\bigg(\! \frac{\vr}{v} \!\bigg)^{2}\frac{\partial^{2}g}{\partial \vr^{2}}+\frac{2\vr\vt}{v^{2}}\frac{\partial^{2}g}{\partial \vt\partial \vr}
  \nonumber
  \\
  &+\bigg(\! \frac{\vt}{v} \!\bigg)^{2}\frac{\partial^{2}g}{\partial \vt^{2}}\bigg] ,
  \\
  \dvPerpSqAvrLoc=&\,4\pi G^{2}\mb\ln\Lambda\bigg[\bigg(\!\frac{\vt}{v}\!\bigg)^{2}\frac{\partial^{2}g}{\partial \vr^{2}}-\frac{2\vr\vt}{v^{2}}\frac{\partial^{2}g}{\partial \vt \partial \vr}
  \nonumber
  \\
  &+\bigg(\! \frac{\vr}{v} \!\bigg)^{2}\frac{\partial^{2}g}{\partial \vt^{2}}+\frac{1}{\vt}\frac{\partial g}{\partial \vt}\bigg] ,
\end{align}
\end{subequations}
where $\Delta v_{\parallel}$ and $\Delta v_{\perp}$ stand respectively
for the local velocity deflections along and perpendicular
to the star's motion. For a fully isotropic cluster,
those reduce to the already known formulae ~\citep[see, e.g.\@, equation~L.25 in][]{Binney2008}, as detailed in \S\ref{sec:iso_coeffs}.

The next step of the calculation is to explicitly compute
all the gradients of the Rosenbluth potentials that appear
in the r.h.s. of equation~\eqref{eq:generic_grad}.
In~\S\ref{sec:Rosenbluth}, owing to an appropriate change of variables,
we rewrite these gradients
as simple three-dimensional integrals over velocity space.
Equation~\eqref{eq:generic_grad} used in conjunction
with equations~\eqref{eq:gh_frame}--\eqref{eq:calc_grad_g_III}
are key results of the present work.
Importantly, these final expressions do not involve
any velocity denominator nor any gradient of the cluster's DF.
In~\S\ref{sec:Chavanis4D}, we also check these new expressions
by deriving them independently from the homogeneous Landau equation.

\subsection{Orbit-average and secular evolution}
\label{subsec:orbit_averaging}

Because
the specific energy and angular momentum
of the test star, $(E,L)$,
are simple functions of $(r,\vr,\vt)$,
it is straightforward to compute
the local diffusion coefficients
in $(E,L)$ from the local velocity diffusion coefficients,
as detailed in~\S\ref{subsec:vel_to_energy}.
These local coefficients are then orbit-averaged
over the unperturbed motion of the test star~\citep[see, e.g.\@, \S{7.4.2} in][]{Binney2008}
by writing
\begin{equation}
D_{E} = \frac{1}{T} \!\int_{0}^{T}\!\! \rd t \, \langle \Delta E \rangle ,
\label{eq:orbit_avg}
\end{equation}
with $T$ the test star's radial period,
and similarly for the other diffusion coefficients.
In equation~\eqref{eq:orbit_avg}, we emphasise
that the integrand, $\langle \Delta E \rangle$,
is evaluated in $(r (t) , \vr (t) , \vt (t))$
as one follows the test star's orbit.
We defer to~\S\ref{subsec:PlummerEffAnomaly}
the presentation of an explicit and numerically stable scheme
to perform this orbit average.

Finally, we rewrite the diffusion coefficients in action space,
namely $\bJ = (\Jr,L)$ with $\Jr$ the radial action,
following~\S\ref{subsec:change_var}.
Rather than focusing on the relaxation of a single test star,
we can treat the whole globular cluster
as a large collection of test stars.
The cluster's orbital distribution is governed
by the reduced DF,
\begin{equation}
F (\bJ) = 2 L \, \Ftot (\bJ) ,
\label{def_Fred}
\end{equation}
which is proportional to the density of stars in $\bJ$-space.
The long-term evolution of $F (\bJ)$ follows from
the Fokker--Planck equation~\citep[see, e.g.\@, \S{7.4} in][]{Binney2008}
which reads here
\begin{align}
\label{def_FP}
\frac{\p F (\bJ)}{\p t} {} &= - \frac{\p }{\p \bJ} \cdot \bF (\bJ)
\\
{} & = - \frac{\p }{\p \bJ} \cdot \bigg[ \bD_{1} (\bJ) \, F (\bJ)- \frac{1}{2} \frac{\p }{\p \bJ} \!\cdot\! \bigg[ \bD_{2} (\bJ) \, F (\bJ) \bigg] \bigg] ,\nonumber
\end{align}
where $\bF (\bJ)$ is the diffusion flux in action space,
and the first- and second-order diffusion coefficients read
\begin{equation}
\bD_{1} (\bJ) =
\begin{pmatrix}
D_{\Jr}
\\
D_{L}
\end{pmatrix}
, \quad
\bD_{2} (\bJ) =
\begin{pmatrix}
D_{\Jr\Jr} & D_{\Jr L}
\\
D_{\Jr L} & D_{L L}
\end{pmatrix} .
\label{def_D1_D2}
\end{equation}
The relaxation rate $\p F / \p t$ in equation~\eqref{def_FP}
then specifies 
the dynamical evolution of any orbit-averaged quantity within the system.

\section{Application to Plummer spheres}
\label{sec:Application}

So far, our derivations apply to any spherically symmetric,
non-rotating anisotropic globular cluster.
Hereafter, we focus on the Plummer sphere
to provide a quantitative assessment
of the impact of anisotropy on secular evolution.

\subsection{The Plummer model}
\label{sec:Plummer}

 The Plummer potential reads
\begin{equation}
\psi (r) = - \frac{G M}{\sqrt{b^{2} + r^{2}}} ,
\label{Plummer_potential}
\end{equation}
with $M$ the cluster's total mass
and $b$ its scale radius.
In the following, for the sake of simplicity,
we use physical units so that
${ G = M = b = 1 }$,
and consider clusters
composed of $N = 10^{5}$ stars.
Following~\cite{Giersz+1994},
we also set the value of the Coulomb logarithm
to $\ln \Lambda = \ln (0.11 \, N)$.

In order to compute orbit-averaged diffusion coefficients
in the case of a Plummer potential,
we rewrite equation~\eqref{eq:orbit_avg} as
\begin{equation}
D_{E} = \frac{2}{ T}\!\int_{-1}^{1} \!\rd u \, \Theta (u) \, \langle \Delta E \rangle .
\label{eq:avgD_u}
\end{equation}
Here, $u$ stands for an explicit effective anomaly
whose weight function $\Theta (u)$ is always well-defined,
numerically stable, and explicit, as detailed in~\S\ref{subsec:PlummerEffAnomaly}.
The same anomaly is also used to rewrite the orbit-averaged
diffusion coefficients in $(\Jr , L)$-space, see~\S\ref{subsec:change_var}.

\begin{figure}
\centering
\includegraphics[width=0.45 \textwidth]{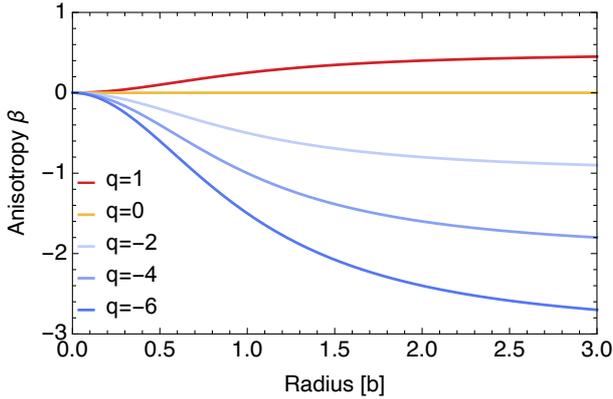}
\caption{Anisotropy parameter $\beta(r)$
of the considered equilibrium DFs (\S\ref{sec:DF})
as a function of the radius in units $b=1$.
Here, $q = 0$ stands for the isotropic equilibrium
while $q < 0$ (resp.\  $q > 0$) are tangentially (resp.\ radially)
anisotropic equilibria.
}
\label{fig:beta}
\end{figure}

In order to vary the cluster's velocity anisotropy,
we consider the same series of equilibria as in~\cite{Breen+2017}.
As detailed in~\S\ref{sec:DF},
the considered DF depends on a parameter $q$
so that its anisotropy parameter $\beta$ varies as
\begin{equation}
\beta(r) = 1-\frac{\sigma_{\rt}^2}{2 \sigma_{\rr}^2} = \frac{q}{2} \frac{r^2}{1+r^2} ,
\label{eq:beta}
\end{equation}
with $\sigma_{\rt}$ and $\sigma_{\rr}$
the local tangential and radial velocity dispersions.
As illustrated in Fig.~\ref{fig:beta},
the case $q = 0$ corresponds to the isotropic equilibrium,
while $q < 0$ (resp.\ $q > 0$) corresponds to tangentially
{(resp.\ radially)} anisotropic equilibria.
Given the radial dependence of equation~\eqref{eq:beta},
we also point out that, for a fixed value of $q$,
the anisotropy is the largest in the cluster's outskirts.

\subsection{Contraction and isotropisation}
\label{sec:Contraction}

\begin{figure}
\centering
  \includegraphics[width=0.45 \textwidth]{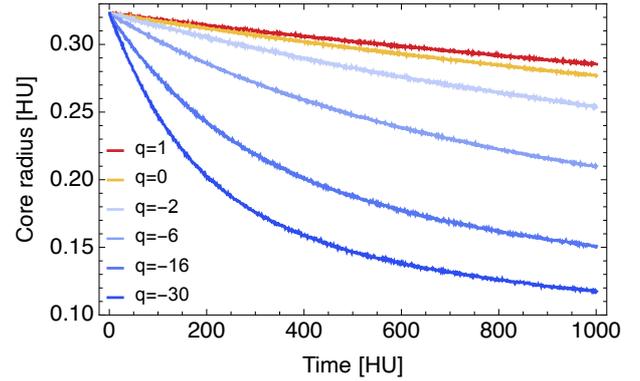}
\caption{Ensemble-averaged evolution of the cluster's core radius as a function of time
from direct $N$-body simulations,
as one varies the anisotropy parameter $q$.
We refer to~\S\ref{sec:NBODY} for the details
of the numerical setup
and the definition of the H\'enon units ($\rHU$).
Larger negative values of $q$,
i.e.\ more tangentially anisotropic clusters,
are unambiguously associated with 
a faster initial evolution.
}
\label{fig:Rc}
\end{figure}

Our goal is to investigate the impact of anisotropy
on the cluster's relaxation.
Using direct $N$-body simulations, detailed in~\S\ref{sec:NBODY},
we illustrate this dependence in Fig.~\ref{fig:Rc}
with the evolution of the cluster's core radius
as one varies $q$.
As already pointed out in~\citet{Breen+2017} (fig.~{4} therein),
the more tangentially anisotropic the cluster,
the faster its initial contraction.

\begin{figure}
\centering
\includegraphics[width=0.45 \textwidth]{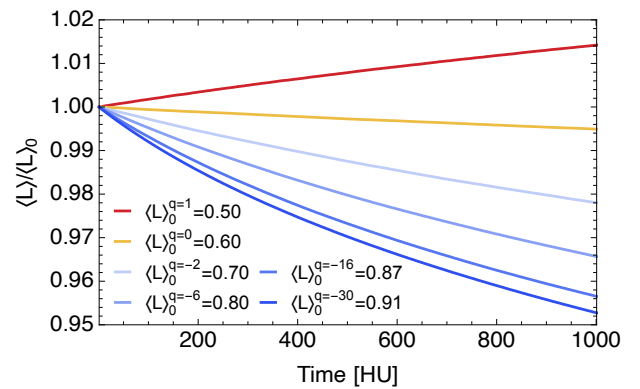}
\caption{Ensemble-averaged evolution of the mean angular momentum norm,
$\langle L \rangle $ (rescaled by its initial value $\langle L \rangle_{0}$)
as one varies the anisotropy parameter $q$,
using the same simulations as in Fig.~\ref{fig:Rc}.
Clusters isotropise throughout their relaxation,
i.e.\ $\langle L \rangle$ increases for radially anisotropic clusters
and decreases for tangentially anisotropic ones.
In addition, for $q < 0$, the stronger the tangential anisotropy,
the faster the initial isotropisation.
}
\label{fig:meanL}
\end{figure}

Using the same simulations,
we also investigate the time evolution
of the clusters' average angular momentum modulus
\begin{equation}
\langle L \rangle =  \frac{(2\pi)^3}{M} \sint \rd \bJ \, L \, F(\bJ) ,
\label{eq:meanL}
\end{equation}
as illustrated in Fig.~\ref{fig:meanL}.
Similarly to fig.~{7} of~\cite{Breen+2017},
we recover here that the clusters' relaxations drive them
towards more isotropic distribution.
Indeed, radially anisotropic clusters (i.e.\ $q > 0$)
see their average angular momentum grow,
i.e.\ orbits become on average more circular,
while tangentially anisotropic clusters (i.e.\ $q < 0$)
see their average angular momentum decrease,
i.e.\ orbits become more radial.
Finally, as in Fig.~\ref{fig:Rc},
we recover that the more tangentially anisotropic a cluster,
the faster its initial isotropisation.

We are now in a position to assess how well
the anisotropic NR theory from~\S\ref{sec:NR}
can predict these trends.

\subsection{Comparing NR and $N$-body evolution}
\label{sec:Relaxation}

In the limit where nearby deflections drive the cluster's relaxation,
its long-term evolution is governed by the Fokker--Planck equation~\eqref{def_FP},
which predicts the rate of change $\p F / \p t$ in action space.

\begin{figure*}
\centering
 \includegraphics[width=1.0 \textwidth]{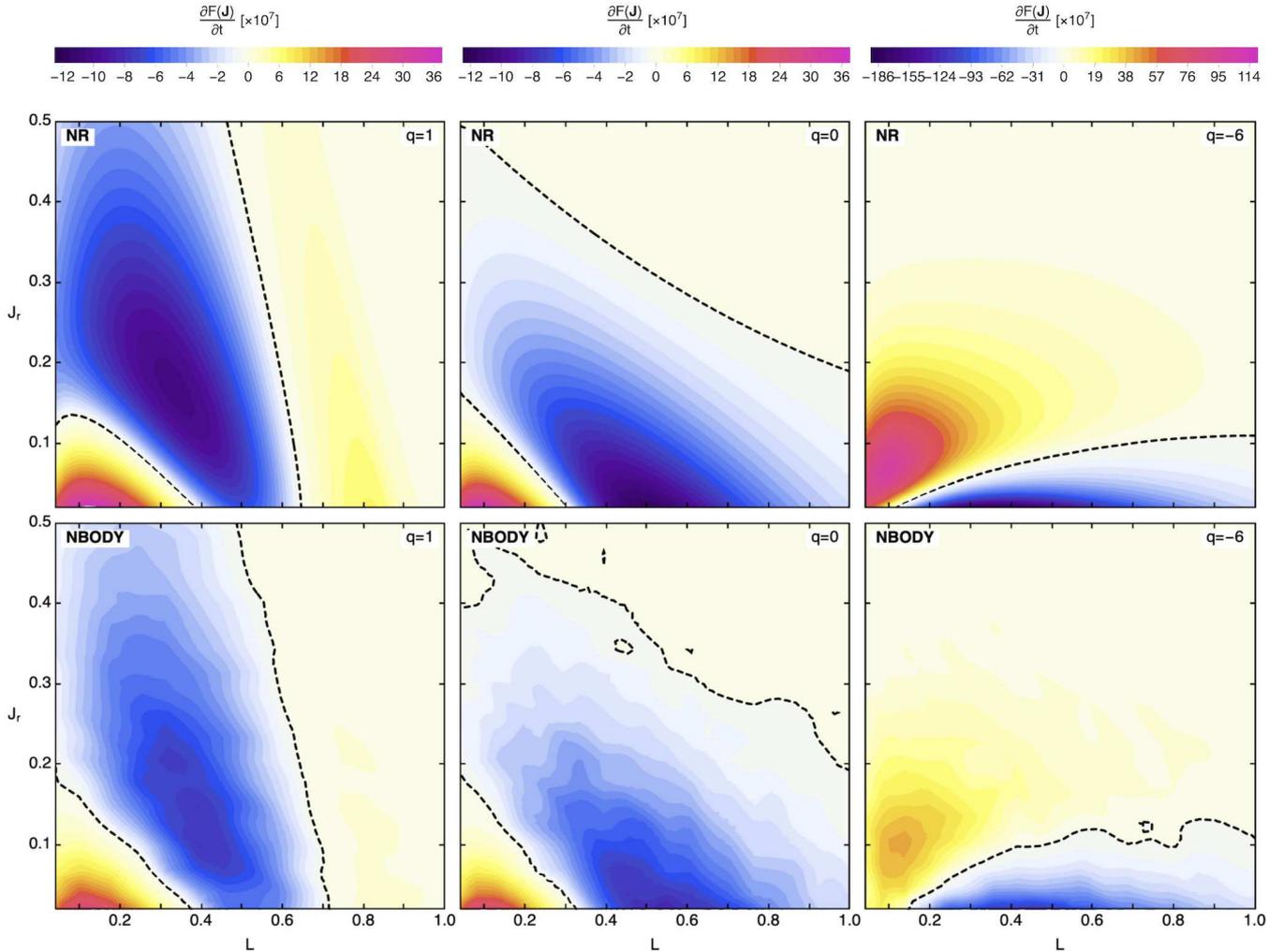}
\caption{Illustration of the local relaxation rate, $\p F / \p t$,
for various values of the anisotropy parameter $q$ (left to right),
as predicted by the anisotropic NR prediction (top, \S\ref{subsec:aniso_diff_coeffs})
and measured in direct numerical simulations (bottom, \S\ref{sec:NBODY}).
There is a qualitative agreement between the NR predictions
and the $N$-body measurement,
up to an overall prefactor depending on the level of anisotropy $q$
(see also Figs.~\ref{fig:Ratio_NR_NBODY} and~\ref{fig:dFdt_NR_NBODY_appendix}).
}
\label{fig:dFdt_NR_NBODY}
\end{figure*}

In Fig.~\ref{fig:dFdt_NR_NBODY},
we compare the contours of $\p F / \p t$
as predicted by the anisotropic NR theory
from equation~\eqref{eq:generic_grad}
with those measured in $N$-body simulations,
for various initial velocity anisotropies\footnote{See~\S\ref{sec:more_aniso} for even more anisotropic distributions.}.
In Fig.~\ref{fig:dFdt_NR_NBODY}, it is remarkable
that the NR maps
and the $N$-body measurements are so similar,
up to an overall prefactor which appears to weakly
depend on the considered actions.
This prefactor reflects the fact that NR theory
poorly accounts for far-away encounters,
that only resonant relaxation captures~\citep[see, e.g.\@,][]{Fouvry+2021}.

\begin{figure}
\centering
\includegraphics[width=0.45 \textwidth]{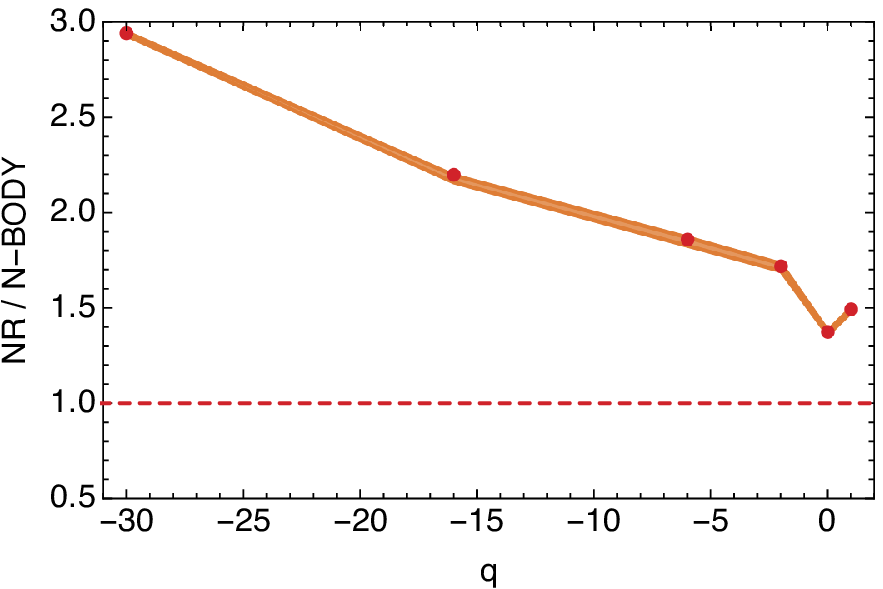}
\caption{Ratio of the diffusion rate of the NR theory
and $N$-body predictions (equation~\ref{eq:Ratio_NR_NBODY})
as a function of the cluster's anisotropy parametrised by $q$.
The dots corresponds to anisotropies for which $N$-body
simulations were performed,
and the contours correspond to the 16\% and 84\%
level lines over the available realisations.
For the isotropic cluster ($q=0$), the NR theory overestimates
the diffusion rate by a factor $\sim 1.4$,
which worsens as the cluster becomes more and more anisotropic.
}
\label{fig:Ratio_NR_NBODY}
\end{figure}

In order to better quantify the overall amount by which the NR
prediction overestimates the $N$-body prediction,
we define the average ratio
\begin{equation}
\frac{\text{NR}}{N\text{-BODY}} = \frac{\displaystyle{\sint \rd \bJ \, F(\bJ)\, |\p F / \p t|_{\text{NR}}}}{\displaystyle{\sint \rd \bJ \, F(\bJ)\, |\p F / \p t|_{N\text{-BODY}}}} ,
\label{eq:Ratio_NR_NBODY}
\end{equation}
where the rates of change, $\p F / \p t$,
are inferred from Fig.~\ref{fig:dFdt_NR_NBODY}.
In Fig.~\ref{fig:Ratio_NR_NBODY}, we present the dependence
of this ratio as a function of the cluster's anisotropies.
In that figure, we find that for an isotropic velocity distribution,
i.e.\ $q=0$, NR overestimates the $N$-body measurement
by a factor $\sim 1.4$.
This is fully compatible
with the previous measurements
from~\cite{Theuns1996} and~\cite{Fouvry+2021}
that observed ratios of order $\sim 1.5-2$
respectively in isotropic King spheres
and isotropic isochrone clusters.
Interestingly, in Fig.~\ref{fig:Ratio_NR_NBODY},
we recover that the ratio from equation~\eqref{eq:Ratio_NR_NBODY}
worsens as the cluster get more tangentially anisotropic:
for $q = - 30$, the NR theory overestimates
the diffusion rate by a factor $\sim 3$.

\begin{figure*}
\centering
\includegraphics[width=1.0 \textwidth]{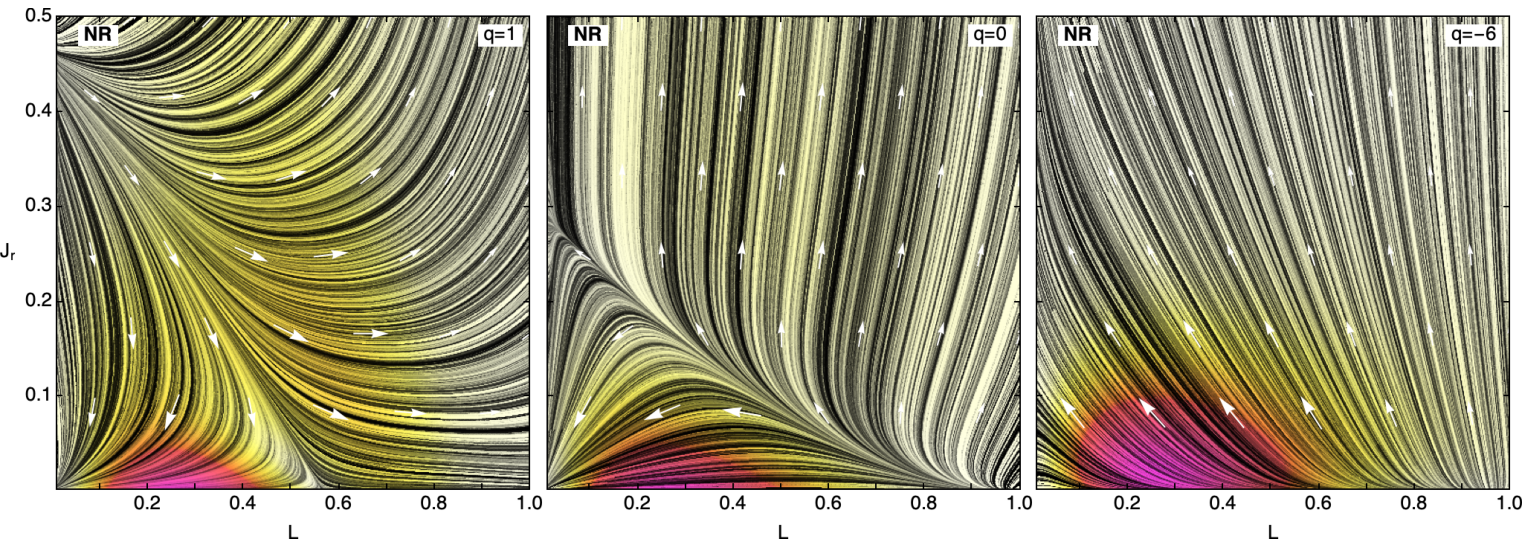}
\caption{Illustration of the field lines of the diffusion flux, $\bF (\bJ)$
(see equation~\ref{def_FP}), as predicted by the NR theory
for various anisotropies.
The arrows give the average direction along which orbits
flow in action space,
while red colors are associated with larger flux amplitudes.
For anisotropic clusters, i.e.\ $q \neq 0$,
these flows reflect the  expected redistribution of orbits towards a more isotropic distribution.
}
\label{fig:flux}
\end{figure*}

In Fig.~\ref{fig:flux}, we provide an alternative
representation of the diffusion predicted by NR.
In that figure, we represent the field lines
of the diffusion flux sourced by equation~\eqref{def_FP},
i.e. the direction along with orbits flow in action space.
Here, we recover that the NR diffusion flux reshuffles the system
towards a more isotropic distribution:
radially anisotropic clusters see their orbits diffuse toward
more circular orbits,
while tangentially anisotropic clusters see their orbits
become more radial on average.

\begin{figure}
\centering
\includegraphics[width=0.45 \textwidth]{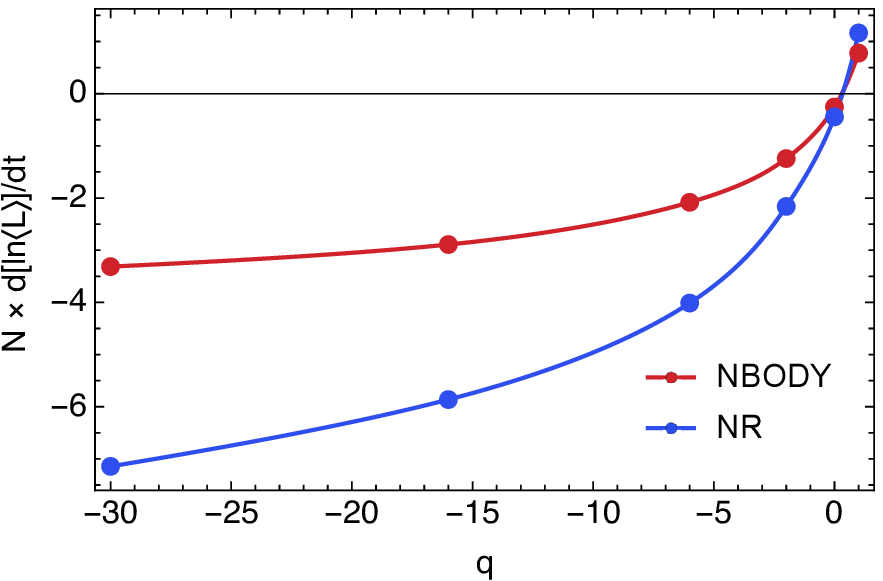}
\caption{Initial value of $\rd \ln \langle L \rangle / \rd t$
as predicted by the NR theory (in blue) and measured in $N$-body simulations (in red).
Anisotropic clusters generically isotropise,
i.e.\ $\rd \langle L \rangle / \rd t > 0$ (resp.\ $< 0$)
in radially (resp.\ tangentially) anisotropic clusters.
}
\label{fig:slopeMeanL}
\end{figure}

Building upon equation~\eqref{eq:meanL},
we can further quantify this isotropisation 
by estimating the initial time variation of the total
angular momentum norm within the sphere via
\begin{equation}
\frac{\rd \langle L \rangle}{\rd t} = \frac{(2 \pi)^{3}}{M} \sint \rd \bJ \, L \, \frac{\p F}{\p t} .
\label{init_slope}
\end{equation}
In Fig.~\ref{fig:slopeMeanL}, we compare $\rd \langle L \rangle / \rd t$
as predicted by the NR theory (using Fig.~\ref{fig:dFdt_NR_NBODY}) 
and as measured in the $N-$body simulations
(using Fig.~\ref{fig:meanL}).
In that figure, we confirm once again that anisotropic
clusters tend to isotropise during their relaxation.
Indeed, one finds $\rd \langle L \rangle / \rd t > 0$
for radially anisotropic clusters ($q > 0$), i.e.\ orbits tend
to get more circular,
while one has $\rd\langle L \rangle / \rd t < 0$
for tangentially anisotropic clusters ($q < 0$),
i.e.\ orbits become more radial.
Similarly to Fig.~\ref{fig:Ratio_NR_NBODY},
we also recover that the NR prediction systematically overestimates
the $N$-body measurement by a factor that grows
as the cluster's initial anisotropy increases.

\section{Discussion}
\label{sec:discussion}

\subsection{Pseudo-Isotropic diffusion}
\label{subsec:PIso}

The anisotropic diffusion coefficients involve
three-dimensional integrals (see equations~\ref{eq:gh_frame}--\ref{eq:calc_grad_g_III}).
This is numerically more demanding than the isotropic ones
which involve one-dimensional integrals (see equation~\ref{eq:1Dintegrals}).
In the view of benefiting from these simpler expressions,
\cite{Cohn1979} introduced the concept of a locally isotropised DF.

Following equation~{(16)} of~\cite{Cohn1979}~\citep[see also equation~{4.81} in][]{Binney2008},
we introduce the pseudo-isotropic (P-Iso) DF\footnote{Equation~\eqref{def_PIso}
follows from equation~{(16)}
of~\cite{Cohn1979} via the change of variable
${ R = \sin^{2} x \, \Rmax }$,
with $R$ and $\Rmax$ defined in~\cite{Cohn1979}.}
\begin{equation}
\Ftot^{\text{P-Iso}}(r,E) = \!\int_0^{\frac{\pi}{2}} \!\! \rd x \, \sin x \, \Ftot\big(E, \sin x \, L_{\max} \big),
\label{def_PIso}
\end{equation}
with $\Lmax (r , E) = \sqrt{2 r^{2} (E - \psi(r))}$
the maximum angular momentum possible
for a bound orbit of energy $E$ going through the radius $r$.
Importantly, following this local average
the pseudo-isotropic DF, $\Ftot^{\text{P-Iso}}$,
only depends on the energy $E$.

\begin{figure}
\centering
\includegraphics[width=0.5 \textwidth]{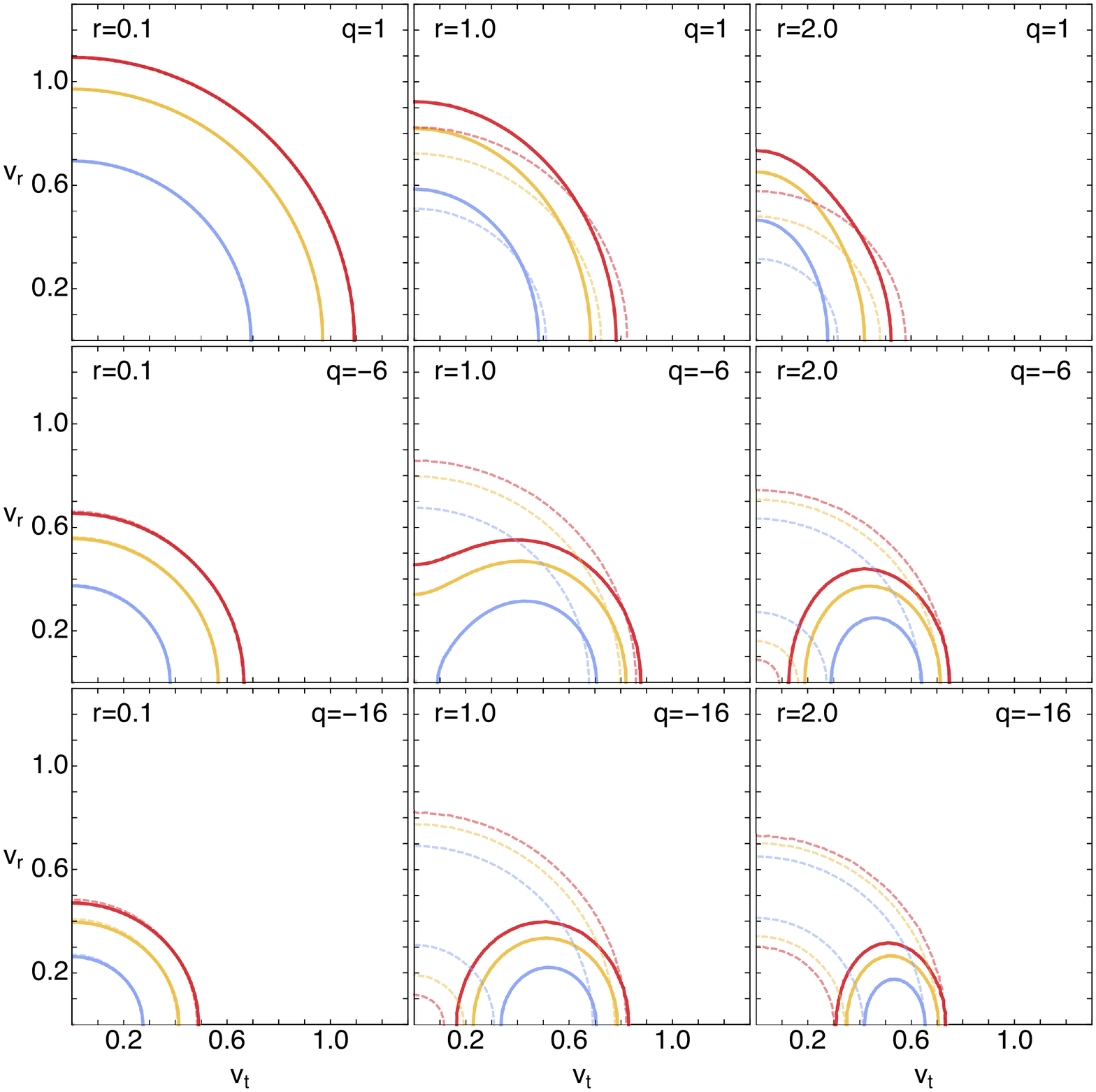}
\caption{Contour levels
of the anisotropic DF (full lines)
and pseudo-isotropic DF (dashed lines)
in the ${ (\vt,\vr) }$-space,
for various radii (left to right)
and various anisotropies (top to bottom).
Contours levels correspond to 50\% (blue),
20\% (orange) and 10\% (red)
of the DF's maximum at the considered radius.
The closer to the cluster's centre,
the weaker the anisotropy,
and therefore the better the match between the two DFs.}
\label{fig:DF_NR_PIso}
\end{figure}

In Fig.~\ref{fig:DF_NR_PIso},
we compare the cluster's anisotropic and pseudo-isotropic DFs
for various radii and various anisotropies.
As already highlighted in Fig.~\ref{fig:beta},
for a fixed value of $q$,
as one moves closer to the cluster's centre,
the local anisotropy diminishes
so that the anisotropic and pseudo-isotropic DFs
closely follow one another.
For a fixed radius $r$, as the anisotropy parameter $q$
gets away from $q=0$,
the local anisotropy increases,
hence increasing the differences between the two DFs.

\begin{figure*}
\centering
\includegraphics[width=1.0 \textwidth]{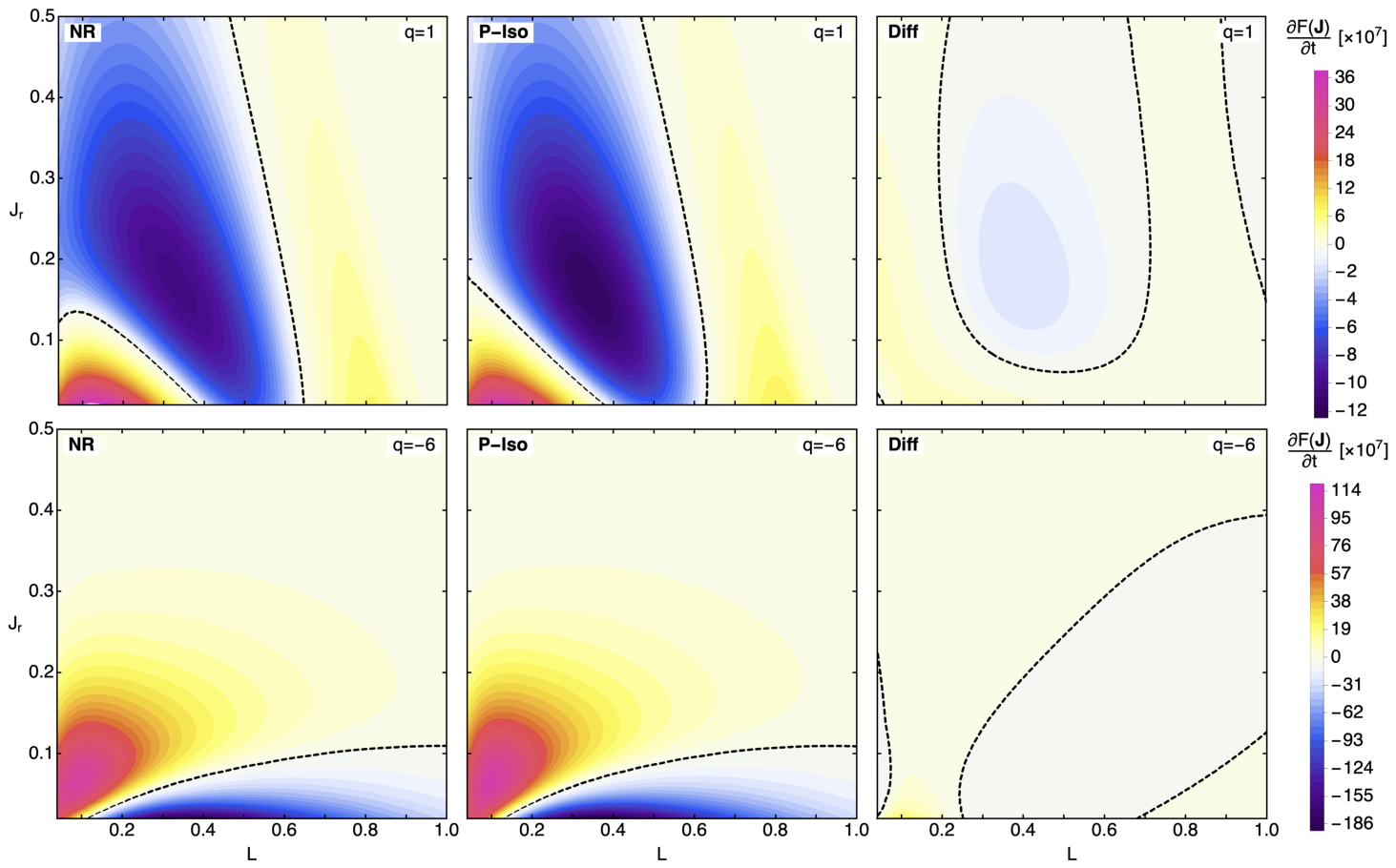}
\caption{Illustration of the relaxation rate, $\p F / \p t$,
for two values of the anisotropy parameter $q$ (top and bottom),
as predicted by the fully anisotropic NR diffusion coefficients (left),
and the pseudo-isotropic ones (middle). The right panel
illustrates the difference (``NR minus P-Iso'').
For the level of anisotropy considered here,
the difference between the two predictions
is found to be, at most, $\sim 5\%$.
}
\label{fig:dFdt_Aniso_PIso}
\end{figure*}

Once the pseudo-isotropic DF is known,
it can straightforwardly be used in equation~\eqref{eq:iso_coeff}
to estimate the velocity diffusion coefficients
via (rapid) one-dimensional integrals (equation~\ref{eq:1Dintegrals}).
This is what we present in Fig.~\ref{fig:dFdt_Aniso_PIso},
where we compare the contours of $\p F / \p t$
as predicted by the fully anisotropic diffusion coefficients
(computed via equation~\ref{eq:generic_grad})
and their pseudo-isotropic analogs
(computed via equation~\ref{eq:iso_coeff}).
In that figure, we note that, for the anisotropies considered here,
the differences between the two maps are minor.
This follows in fact from Fig.~\ref{fig:beta},
where we noted that as one moves closer to the cluster's core,
the anisotropy gets reduced,
hence the similitude of the two maps reported in Fig.~\ref{fig:dFdt_Aniso_PIso}
which focus on the cluster's central region.
We reach
the same conclusion in~\S\ref{sec:pseudo_aniso}
by noting that the local velocity deflections
accumulated along a test star's motion
in the cluster's core
only marginally differ
between the anisotropic and pseudo-isotropic predictions
(see Fig.~\ref{fig:local_vel_contrib}).

\begin{figure}
\centering
\includegraphics[width=0.45 \textwidth]{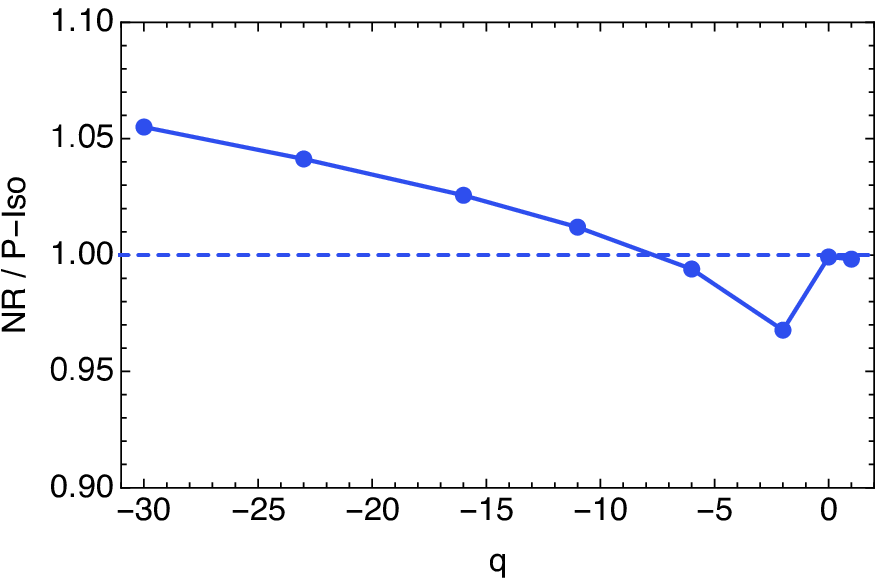}
\caption{Ratio of the diffusion rate of the NR prediction
and the P-Iso one, as defined in equation~\eqref{eq:Ratio_NR_PIso}.
For an isotropic cluster ($q=0$) both predictions
are, naturally, in agreement,
and they start to differ as the cluster
becomes more and more anisotropic.
}
\label{fig:Ratio_NR_PIso}
\end{figure}

In order to better compare these two predictions,
following equation~\eqref{eq:Ratio_NR_NBODY},
we compute the respective ratio of the NR
and P-Iso predictions through
\begin{equation}
\frac{\text{NR}}{\text{P-Iso}} = \frac{\displaystyle{\sint \rd \bJ \, F(\bJ) \,|\partial F/\partial t|_{\text{NR}}}}{\displaystyle{\sint \rd \bJ \, F(\bJ)\, |\partial F/\partial t|_{\text{P-Iso}}}} ,
\label{eq:Ratio_NR_PIso}
\end{equation}
which is represented in Fig.~\ref{fig:Ratio_NR_PIso}.
In that figure, we recover that for the anisotropy parameters $q$
considered here, the two maps typically differ, at most, by $\sim 5\%$.
As expected, as one increases the cluster's anisotropy,
the mismatch between the two predictions increases.
Finally, we point out that this similitude between NR and P-Iso
is a pleasant numerical news.
Indeed, rather than having to go through
the three-dimensional NR integrals from equations~\eqref{eq:gh_frame}--\eqref{eq:calc_grad_g_III},
the P-Iso prediction requires, in essence, two-dimensional integrals
by computing first the pseudo-isotropic DF from equation~\eqref{def_PIso}
and subsequently the associated isotropic
diffusion coefficients from equation~\eqref{eq:1Dintegrals}.

\subsection{Conclusion}
\label{sec:conclusion}

We tailored Chandrasekhar's NR theory
to compute the local velocity diffusion coefficients
in non-rotating anisotropic spherical clusters.
We implemented explicitly these expressions,
now distributed through a publicly available code.

We subsequently applied the present anisotropic NR theory
to  a series of anisotropic Plummer spheres.
Building upon~\cite{Breen+2017},
we reached two main conclusions.
First, the NR prediction
matches qualitatively direct $N$-body measurements
(see Fig.~\ref{fig:dFdt_NR_NBODY})
up to an overall prefactor $\sim 1.4 - 2$
that worsens as the initial anisotropy increases
(see Fig.~\ref{fig:Ratio_NR_NBODY}).
This match between theory and simulations
shows that NR captures the more rapid compression of tangentially anisotropic globular cluster compared to radially anisotropic ones (see Fig.~\ref{fig:Rc}),
though the incorrect prefactor suggests 
that NR misestimates the contributions
from large-scale encounters.
Second, we pointed out that NR
also drives initially an isotropisation of the clusters
(see, e.g.\@, Fig.~\ref{fig:flux}).

Finally, following~\cite{Cohn1979},
we investigated the errors introduced
by locally isotropising the DF of the perturbers
(see equation~\ref{def_PIso}),
an approach that we coined ``pseudo-isotropic''.
For the class of anisotropic clusters considered,
we emphasised that the limited extent of anisotropy
in the inner regions (see Fig.~\ref{fig:DF_NR_PIso})
led to differences of order $\lesssim 5\%$ with respect to
the fully anisotropic calculation.
As such, for the clusters considered here,
we confirmed that the anisotropy of the perturbers' DF,
via $\Ftot(\br,\bvpp)$ in equation~\eqref{eq:def_gh},
plays a much less important role than the anisotropy
in the test particles' DF, via $F(\bJ)$ in equation~\eqref{def_FP}.

\subsection{Perspectives}
\label{sec:perspectives}

Having computed $\p F / \p t$
from kinetic theory (Fig.~\ref{fig:dFdt_NR_NBODY}),
we could in principle  predict the initial time evolution
of more traditional quantities such
as the anisotropy parameter $\beta (r)$ (Fig.~\ref{fig:beta})
or the core radius $\Rc (t)$ (Fig.~\ref{fig:Rc})
at the cost of accounting appropriately
for the self-consistent update of the cluster's mean potential.
Ultimately, following for example~\cite{Vasiliev2015},
one could also hope to integrate self-consistently
the time evolution of the whole cluster
as driven by the present NR theory.
Figure~\ref{fig:flux} emphasised
that local deflections
naturally tend to isotropise the cluster's DF.
Given this increased isotropy,
one may expect that the pseudo-isotropic prescription
from equation~\eqref{def_PIso} will become
 more relevant as the relaxation occurs.
Of course this would  deserve to be investigated in more detail, following for example fig.~{7} of~\cite{Breen+2017}.

In~\S\ref{sec:Application},
we restricted  our analysis to Plummer potentials.
Nevertheless the generic derivation presented
in~\S\ref{sec:NR} should translate to any ``reasonable'' core potential,
provided one has access to its distribution function,
e.g.\@, following the method from~\cite{Dejonghe1987}.
It would also be of interest
to investigate truncated, cuspy, or even rotating
spheres, since the clusters' orbital structure
impacts both their linear and long-term responses.
In order to alleviate some of the numerical challenges,
it would be worthwhile to find an efficient way
of carrying out the orbit-averages
using numerically stable effective anomalies
(as in~\S\ref{subsec:PlummerEffAnomaly})
for each such potential.

As illustrated in Fig.~\ref{fig:Ratio_NR_NBODY},
the NR theory and the $N$-body measurements
still present an overall multiplicative  discrepancy.
It most probably arises from the fact that 
NR does not capture accurately
the contribution from far-away  encounters,
collective effects, and non-local resonances.
Following~\cite{Fouvry+2021}, it would be of interest
to investigate the resonant relaxation (RR)
of spherical clusters with various levels of anisotropy.
This should ultimately pave the way to predict ab initio
the effective Coulomb logarithm $\ln \Lambda$ in equation~\eqref{eq:generic_D}.
This will be the topic of future work.

\subsection*{Data Distribution}
The  data underlying this article 
is available through reasonable request to the author.
The code for the  anisotropic NR diffusion coefficient
is available at the following URL: \href{https://github.com/KerwannTEP/CAT}{https://github.com/KerwannTEP/CAT}.

\section*{Acknowledgements}

This work is partially supported by grant Segal ANR-19-CE31-0017
of the French Agence Nationale de la Recherche,
and by the Idex Sorbonne Universit\'e.
We are grateful to M.\ Roule and M.\ Petersen
for numerous suggestions during the completion of this work.
We thank St\'ephane Rouberol for the smooth running of the
Infinity cluster, where the simulations were performed.

\appendix

\section{Local diffusion coefficients}
\label{sec:LocalDiff}

The local diffusion coefficients, $\langle \Delta v_{i} \rangle$
and $\langle \Delta v_{i} \Delta v_{j} \rangle$,
are generically
given by equation~\eqref{eq:generic_D},
where $1 \leq i,j \leq 3$ are associated
with an arbitrary frame.
In order to compute the diffusion coefficients
for a generic anisotropic DF, we proceed
via two consecutive steps.
First, in this Appendix, we generically rewrite the gradients
of the Rosenbluth potentials as gradients
with respect to the radial and tangential velocities $(\vr,\vt)$.
Then, in~\S\ref{sec:Rosenbluth}, we explicitly compute these gradients
by rewriting them as simple three-dimensional integrals
over velocity space.

Assuming that the DF at play is written under the form $\Ftot (r,\vr,\vt)$,
it is natural to aim at expressing the Rosenbluth potentials
as $h(r,\vr,\vt)$ and $g(r,\vr,\vt)$,
and compute their gradients with respect to these same coordinates.
To do so, we first compute in~\S\ref{subsec:arbitraryFrame}
the gradients of $h$ and $g$ in an arbitrary coordinate system.
Then, in~\S\ref{subsec:specialFrame},
we apply these generic expressions to a tailored
frame to ultimately obtain
our main result in equation~\eqref{eq:generic_grad}.

\subsection{Arbitrary frame}
\label{subsec:arbitraryFrame}

\begin{figure}
\centering
\includegraphics[width=0.35 \textwidth]{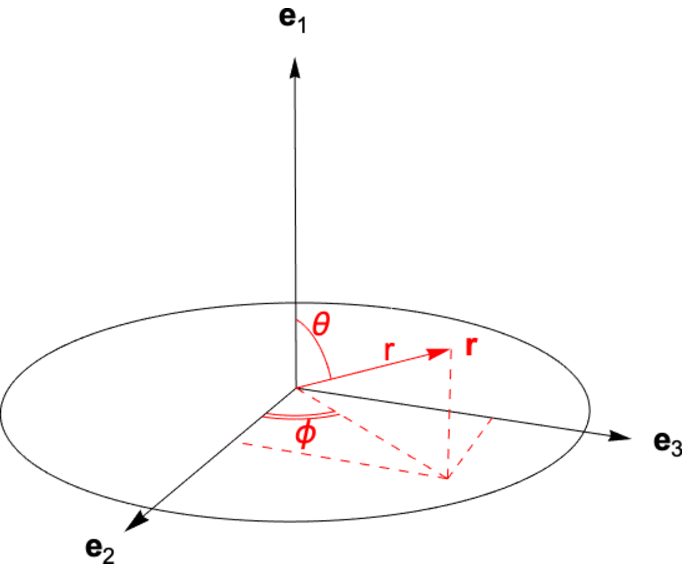}
\caption{Arbitrary frame used to describe the test particle's position,
as in equation~\eqref{br_coordinate}.
}
\label{fig:Coord_syst_1}
\end{figure}

As illustrated in Fig.~\ref{fig:Coord_syst_1},
we denote our initial arbitrary frame
as $(\e1,\e2,\e3)$, with $\e1$ the $z$-axis.
Using standard spherical coordinates,
we parametrise the current position of the test particle,
$\br = \br (r , \theta , \phi)$, as
\begin{equation}
r_{1} = r\cos\theta ,\quad
r_{2} = r\sin\theta\cos\phi , \quad
r_{3} = r\sin\theta\sin\phi .
\label{br_coordinate}
\end{equation}
The test particle's velocity $\bv = (v_{1} , v_{2} , v_{3})$
can then be decomposed into its radial and tangential velocities as
\begin{align}
 \vr &= \bv \cdot\hr =v_{1}\cos\theta\!+\!v_{2}\sin\theta\cos\phi\!+\!v_{3}\sin\theta\sin\phi ,
 \label{eq:vRadTan_coord}
 \\
 \vt^{2} &= v^{2}\!\!-\!\vr^{2} =v^{2}\!-\!(v_{1}\cos\theta\!+\!v_{2}\sin\theta\cos\phi\!+\!v_{3}\sin\theta\sin\phi)^2\! ,
\notag
\end{align}
with the usual notation $\hr = \br/|\br|$.

From equation~\eqref{eq:generic_D},
the diffusion coefficients involve the first-order (resp.\ second-order)
gradient of $h$ (resp.\ $g$).
For $h (\br,\bv) = h (r , \vr , \vt)$,
one simply has
\begin{equation}
\frac{\partial h}{\partial v_{1}}=\frac{\partial h}{\partial \vr}\frac{\partial \vr}{\partial v_{1}} +\frac{\partial h}{\partial \vt}\frac{\partial \vt}{\partial v_{1}}\ ,
\label{eq:generic_grad_h}
\end{equation}
and similarly for $\p h /\p v_{2}$ (resp.\ $\p h / \p v_{3}$)
by the direct replacement $v_{1} \to v_{2}$ (resp.\ $v_{1} \to v_{3}$).
In equation~\eqref{eq:generic_grad_h},
the relevant velocity gradients
follow from equation~\eqref{eq:vRadTan_coord}
and generically read
\begin{subequations}
\label{eq:dvdv1}
\begin{align}
 \displaystyle{\frac{\partial\vr}{\partial v_{1}}} &= \displaystyle{\cos\theta} ,  &\displaystyle{\frac{\partial \vt}{\partial v_{1}}} =& \displaystyle{\frac{v_{1}-\vr\cos\theta}{\vt}},
 \\
 \displaystyle{\frac{\partial\vr}{\partial v_{2}}}&= \displaystyle{\sin\theta \cos\phi} ,  &\displaystyle{\frac{\partial \vt}{\partial v_{2}}} =& \displaystyle{\frac{v_{2}-\vr\sin\theta\cos\phi}{\vt}},
 \\
 \displaystyle{\frac{\partial\vr}{\partial v_{3}}}&= \displaystyle{\sin\theta \sin\phi} ,  &\displaystyle{\frac{\partial \vt}{\partial v_{3}}} =& \displaystyle{\frac{v_{3}-\vr\sin\theta\sin\phi}{\vt}} .
\end{align}
\end{subequations}
Applying the chain rule,
we can similarly compute the second-order gradients
of $g(\br,\bv) = g (r , \vr , \vt)$ via
\begin{align}
\frac{\partial^{2}g}{\partial v_{1}^{2}}&=\frac{\partial^2 \vr}{\partial v_{1}^2}\frac{\partial g}{\partial \vr}+\left(\!\frac{\partial \vr}{\partial v_{1}}\!\right)^{2}\frac{\partial^{2}g}{\partial \vr^{2}}+\frac{\partial \vr}{\partial v_{1}}\frac{\partial \vt}{\partial v_{1}}\frac{\partial^{2}g}{\partial \vr\partial \vt} \notag\\
 &+\frac{\partial^2 \vt}{\partial v_{1}^2}\frac{\partial g}{\partial \vt}+\left(\!\frac{\partial \vt}{\partial v_{1}}\!\right)^{2}\frac{\partial^{2}g}{\partial \vt^{2}}+\frac{\partial \vt}{\partial v_{1}}\frac{\partial \vr}{\partial v_{1}}\frac{\partial^{2}g}{\partial \vt\partial \vr} ,
 \label{eq:generic_grad_g}
\end{align}
where the derivatives $\p^{2} g /\p v_{2}^{2}$
and $\p^{2} g / \p v_{3}^{2}$ are obtained by direct replacements.
In equation~\eqref{eq:generic_grad_g},
the only non-zero second-order velocity gradients
follow once again from equation~\eqref{eq:vRadTan_coord}
and read
\begin{subequations}
\label{eq:d2gdv12_Coeffs_Arbitrary}
\begin{align}
 \frac{\partial^2 \vt}{\partial v_{1}^2}&\!=\!\frac{\cos\theta}{\vt}\left(
  {\sin^{2}\theta-\frac{(v_{1}-\vr\cos\theta)^2}{\vt^2}}
\right) ,\\
\frac{\partial^2 \vt}{\partial v_{2}^2}&\!=\! \frac{1}{\vt}\!\left(\!{(1\!-\!\sin^{2}\theta\cos^{2}\phi)\!-\!\frac{(v_{2}\!-\!\vr\sin\theta\cos\phi)^2}{\vt^2}}\!\right), \\\!
  \frac{\partial^2 \vt}{\partial v_{3}^2}&\!=\!\frac{1}{\vt} \bigg(\! (1\!-\!\sin^{2}\theta\sin^{2}\phi)\!-\!\frac{(v_{3}\!-\!\vr\sin\theta\sin\phi)^2}{\vt^2} \!\bigg).
\end{align}
\end{subequations}

\subsection{Special frame}
\label{subsec:specialFrame}

\begin{figure}
\centering
\includegraphics[width=0.35 \textwidth]{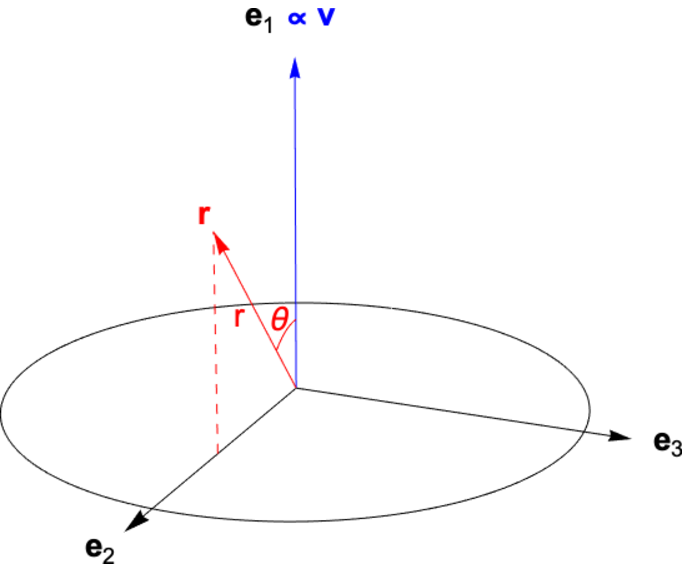}
\caption{Tailored frame used to compute the parallel
and perpendicular local velocity deflections
in equation~\eqref{eq:SpecialCoordSys_LocDiffCoeffs}.
}
\label{fig:Coord_syst_1_spe}
\end{figure}

Having obtained generic expressions
for the needed gradients, we can now evaluate
them in an appropriate frame.
More precisely, as illustrated in Fig.~\ref{fig:Coord_syst_1_spe},
we consider the particular frame
$\e1 = \hv$ and $\phi = 0$.
In that  frame,
the velocities simply reduce
to $v_{1}=v$, $v_{2}=v_{3}=0$.
Similarly, we also have $\cos\theta = \vr / v$,
and $\sin\theta = \vt/v$.

Benefiting from these simple relations,
we can now use equations~\eqref{eq:generic_grad_h} and~\eqref{eq:dvdv1}
to get a simple expression for the first-order gradient
of $h$ as
\begin{equation}
\frac{\partial h}{\partial v_{1}}=\frac{\vr}{v}\frac{\partial h}{\partial \vr}+\frac{\vt}{v}\frac{\partial h}{\partial \vt},
\label{eq:needed_grad_h}
\end{equation}
Similarly, using equations~\eqref{eq:generic_grad_g} and~\eqref{eq:d2gdv12_Coeffs_Arbitrary},
we also obtain a simple expression for the needed
second-order gradients of $g$ via
\begin{subequations}
\label{eq:needed_grad_g}
\begin{align}
\frac{\partial^{2}g}{\partial v_{1}^{2}}  &=\left(\frac{\vr}{v}\right)^{2}\frac{\partial^{2}g}{\partial \vr^{2}}+\frac{2\vr\vt}{v^{2}}\frac{\partial^{2}g}{\partial \vt\partial \vr}+\left(\frac{\vt}{v}\right)^{2}\frac{\partial^{2}g}{\partial \vt^{2}},
\\
\frac{\partial^{2}g}{\partial v_{2}^{2}}  &=\left(\frac{\vt}{v}\right)^{2}\frac{\partial^{2}g}{\partial \vr^{2}}-\frac{2\vr\vt}{v^{2}}\frac{\partial^{2}g}{\partial \vt\partial \vr}+\left(\frac{\vr}{v}\right)^{2}\frac{\partial^{2}g}{\partial \vt^{2}},
\\
\frac{\partial^{2}g}{\partial v_{3}^{2}} &= \frac{1}{\vt}\frac{\partial g}{\partial \vt} .
\end{align}
\end{subequations}
In the particular frame from Fig.~\ref{fig:Coord_syst_1_spe},
one has
$\Delta v_{1}=\dvPar$ and $\Delta v_{2}^{2}+\Delta v_{3}^{2}=\dvPerp^{2}$,
with $\dvPar$ and $\dvPerp$
the velocity deflections parallel and perpendicular
to the star's motion.
This translates into the local velocity diffusion coefficients
\begin{subequations}
\label{eq:SpecialCoordSys_LocDiffCoeffs}
\begin{align}
  \dvParAvrLoc&=\langle\Delta v_{1}\rangle ,
  \\
  \dvParSqAvrLoc&=\langle(\Delta v_{1})^{2}\rangle ,
  \\
  \dvPerpSqAvrLoc &=\langle(\Delta v_{2})^{2}\rangle+\langle(\Delta v_{3})^{2}\rangle .
\end{align}
\end{subequations}
The final step of the calculation is then
to inject the expressions of the gradients
from equations~\eqref{eq:needed_grad_h} and~\eqref{eq:needed_grad_g}
into equation~\eqref{eq:generic_D}.
Overall, this leads to the final result from equation~\eqref{eq:generic_grad}.

One could fear that equation~\eqref{eq:generic_grad}
presents singularities for $v = 0$ and $\vt = 0$.
Fortunately, those are only coordinate singularities.
Indeed the limiting case $v = 0$ only occurs
at pericentre and apocentre of radial orbits, i.e.\ orbits with $\vt = 0$.
In that limit, one has $v=\vr$,
so that $\vr/v=1$ and $\vt/v=0$,
resolving the $1/v$ singularities
in equation~\eqref{eq:generic_grad}.
As for the $1/\vt$ singularity,
we note that $\p g / \p \vt$ vanishes in $\vt = 0$
 -- as shown in \S \ref{subsec:Rosenbltuh_properties} --
so that $\lim_{\vt \to 0} (\p g/\p \vt)/\vt = \p^2 g / \p \vt^2$.
This ensures similarly that there are no $1/\vt$ singularities
in equation~\eqref{eq:generic_grad}.

\section{Isotropic diffusion coefficients}
\label{sec:iso_coeffs}

The anisotropic diffusion coefficients from equation~\eqref{eq:generic_grad}
must reduce to the known isotropic expressions
in the limit of an isotropic DF, $\Ftot = \Ftot (E)$.
We check it in this Appendix.

Assuming that the cluster is isotropic,
we generically have $h (\vr,\vt) = h(v)$
with $v^{2} = \vr^{2} + \vt^{2}$
and similarly for $g(\vr,\vt) = g (v)$.
Focusing back on equation~\eqref{eq:generic_grad},
we can rewrite the derivatives appearing in $\dvParAvrLoc$ as
\begin{equation}
\frac{\p h}{\p \vr} = \frac{\vr}{v} \frac{\p h}{\p v} ,
\quad
\frac{\p h}{\p \vt} = \frac{\vt}{v} \frac{\p h}{\p v} .
\label{eq:AniToIso_d1}
\end{equation}
As required by equation~\eqref{eq:generic_grad},
we obtain
\begin{equation}
\frac{\vr}{v}\frac{\partial h}{\partial \vr}+\frac{\vt}{v}\frac{\partial h}{\partial \vt} = \frac{\p h}{\p v} .
\label{eq:grad_iso_h}
\end{equation}
Similarly, the derivatives involved in  $\dvParSqAvrLoc$
and $\dvPerpSqAvrLoc$ in equation~\eqref{eq:generic_grad}
generically read
\begin{subequations}
\label{eq:AniToIso_d2}  
\begin{align}
\frac{\partial^{2}g}{\partial \vr^{2}} &=\frac{\partial}{\partial \vr}\bigg(\frac{\partial v}{\partial \vr}\frac{\partial g}{\partial v}(v)\bigg) =\frac{\vt^{2}}{v^{3}}\frac{\partial g}{\partial v}+\frac{\vr^{2}}{v^{2}}\frac{\partial^2 g}{\partial v^2},
\\
 \frac{\partial^{2}g}{\partial \vt\partial \vr}&=\frac{\partial}{\partial\vt}\bigg(\frac{\partial v}{\partial \vr}\frac{\partial g}{\partial v}(v)\bigg) \!=\!-\frac{\vr\vt}{v^{3}}\frac{\partial g}{\partial v}+\frac{\vr\vt}{v^{2}}\frac{\partial^2 g}{\partial v^2},\,
\\
\frac{\partial^{2}g}{\partial \vt^{2}}&=\frac{\partial}{\partial \vt}\bigg(\frac{\vt}{v}\frac{\partial g}{\partial v}(v)\bigg)=\frac{\vr^{2}}{v^{3}}\frac{\partial g}{\partial v}+\frac{\vt^{2}}{v^{2}}\frac{\partial^2 g}{\partial v^2},
\\
\frac{\partial g}{\partial \vt}&=\frac{\partial v}{\partial \vt}\frac{\partial g}{\partial v}(v)=\frac{\vt}{v}\frac{\partial g}{\partial v} .
\end{align}
\end{subequations}
Therefore, in the isotropic limit,
we obtain the simplifications
\begin{subequations}
\label{eq:grad_iso_g}
\begin{align}
 & \bigg(\!\!\frac{\vr}{v}\!\!\bigg)^{2} \! \frac{\partial^{2}g}{\partial \vr^{2}}\!+\!\frac{2\vr\vt}{v^{2}}\frac{\partial^{2}g}{\partial \vt\partial\vr}\!+\!\bigg(\!\! \frac{\vt}{v} \!\!\bigg)^{2}\frac{\partial^{2}g}{\partial \vt^{2}} \!=\! \frac{\partial^2 g}{\partial v^2},
 \\
 &  \bigg(\!\!\frac{\vt}{v}\!\!\bigg)^{2}\!\frac{\partial^{2}g}{\partial \vr^{2}}\!-\!\frac{2\vr\vt}{v^{2}}\frac{\partial^{2}g}{\partial\vt\partial \vr}\!+\!\bigg(\!\!\frac{\vr}{v}\!\!\bigg)^{2} \! \frac{\partial^{2}g}{\partial \vt^{2}}\!+\!\frac{1}{\vt}\frac{\partial g}{\partial \vt} \!=\! \frac{2}{v}\frac{\partial g}{\partial v}. 
\end{align}
\end{subequations}
Finally,  plugging equations~\eqref{eq:grad_iso_h} and~\eqref{eq:grad_iso_g}
into the local velocity diffusion coefficients
from equation~\eqref{eq:generic_grad}
yields the local isotropic diffusion coefficients~\citep[][ equation~L.25]{Binney2008} 
\begin{subequations}
\label{eq:iso_coeff}
\begin{align}
\dvParAvrLoc &=4\pi G^2 (m+\mb) \ln \Lambda\, \frac{\partial h}{\partial v} ,
\\
\dvParSqAvrLoc &= 4\pi G^2 \mb \ln \Lambda\,\frac{\partial^2 g}{\partial v^2} ,
\\
\dvPerpSqAvrLoc &=\frac{8\pi G^2 \mb \ln \Lambda}{v}\,\frac{\partial g}{\partial v} . 
\end{align}
\end{subequations}
In practice, the derivatives from equation~\eqref{eq:iso_coeff}
can be evaluated via
\begin{subequations}
\label{eq:grad_iso}
\begin{align}
\frac{\partial h}{\partial v} &=-\frac{4\pi}{v}K_1(r,v) ,
\\
\frac{\partial g}{\partial v} &= \frac{8\pi v}{3 } \bigg(2K_0(r,v)+ 3 K_1(r,v) - K_3(r,v) \bigg),
\\
\frac{\partial^2 g}{\partial v^2} &= \frac{8\pi}{3 } \bigg(K_0(r,v)+K_3(r,v)\bigg),
\end{align}
\end{subequations}
where we defined the integrals
\begin{subequations}
\label{eq:1Dintegrals}
\begin{align}
K_0(r,v) &=\int_{v}^{+\infty}\hspace{-3mm} \rd \vpp\vpp \, \Ftot(r,v'),
\\
K_1(r,v) &= \int_{0}^{v} \rd \vpp\vpp \, \bigg(\frac{\vpp}{v}\bigg) \,\Ftot(r,v'),
\\
K_3(r,v) &= \int_{0}^{v} \rd \vpp\vpp \, \bigg(\frac{\vpp}{v}\bigg)^{3} \,\Ftot(r,v').
\end{align}
\end{subequations}

\section{Gradients of potentials}
\label{sec:Rosenbluth}

In equation~\eqref{eq:generic_grad},
we obtained a generic expression of the anisotropic
diffusion coefficients as functions of the gradients
of the Rosenbluth potentials with respect to $(\vr,\vt)$.
In this Appendix, we obtain explicit expressions
for the needed gradients as simple non-singular  three-dimensional
integrals over velocity space.

\subsection{Derivation of gradient expressions}

The Rosenbluth potentials, $h$ and $g$,
are generically given by equation~\eqref{eq:def_gh}.
Our goal is to pick an appropriate frame
to express them as $h(r,\vr,\vt)$ and $g(r,\vt,\vt)$,
so as to compute their gradients with respect to $\vr$ and $\vt$.

Following Fig.~\ref{fig:Coord_syst_3},
we now fix our frame to be
$\e{z} = \hr$ and $\e{x} = \bvt/\vt$.
\begin{figure}
\centering
\includegraphics[width=0.35\textwidth]{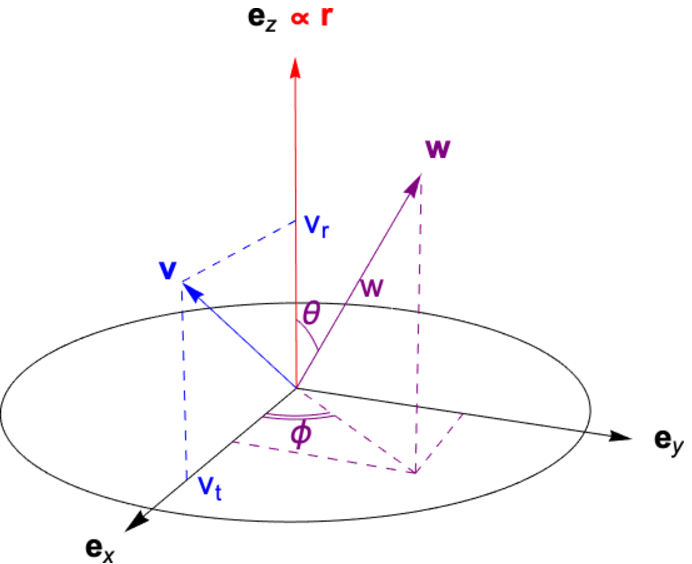}
\caption{Tailored frame used to compute
$h(r,\vr,\vt)$ and $g(r,\vr,\vt)$
as in equation~\eqref{def_bw}.}
\label{fig:Coord_syst_3}
\end{figure}
In that frame, the velocity of the test particle
is simply $\vx = \vt$, $\vy = 0$, $\vz=\vr$.
In the same frame, we can decompose
the velocity difference $\bw = \bv - \bvpp$ as
\begin{equation}
w_{ x} \!=\! w\sin\theta\cos\phi ,
\;
w_{ y} \!=\! w\sin\theta\sin\phi ,
\;
w_{ z} \!=\! w\cos\theta . 
\label{def_bw}
\end{equation}
As a consequence, the velocity
of the background particle, $\bvpp = \bv - \bw$
reads
\begin{equation}
\vppx  \!=\!\vt\! -\! w\sin\!\theta\!\cos\!\phi , \,
  \vppy \!=\! - w\sin\!\theta\!\sin\!\phi , \,
  \vppz  \!=\! \vr \!-\! w\cos\!\theta ,
  \label{eq:chg_ref}
\end{equation}
with the associated radial and tangential velocity decomposition
\begin{subequations}
\label{eq:vrpvtp_def}
\begin{align}
  \vppr & =\vr-w\cos\theta ,
  \\
  \vppt & =\sqrt{(\vt-w\sin\theta\cos\phi)^{2}+(w\sin\theta\sin\phi)^{2} } .
\end{align}
\end{subequations}
The energy and angular momentum $(\Epp,\Lpp)$ of the background particle
are then naturally given by
\begin{equation}
  \Epp  =\psi(r)+\half \vpprSq + \half \vpptSq ,
  \quad
  \Lpp  =r \vppt  .
  \label{eq:Ea_La}
\end{equation}
Within that same frame,
the Rosenbluth potentials from equation~\eqref{fig:Coord_syst_3}
read
\begin{subequations}
\label{eq:gh_frame}
\begin{align}
h (\br , \bv) {} & = \sint \rd w \rd \theta \rd \phi \, w \sin \theta \, \Ftot(r , \bv \!-\! \bw) ,
\\
g (\br , \bv) {} & = \sint \rd w \rd \theta \rd \phi \, w^{3} \sin \theta \, \Ftot (r , \bv \!-\! \bw) .
\end{align}
\end{subequations}
We are now in a position to compute the needed
gradients of the Rosenbluth potentials
present in equation~\eqref{eq:generic_grad}.
Starting from equation~\eqref{eq:def_gh},
we must compute
\begin{subequations}
\label{eq:calc_grad_h}
\begin{align}
\frac{\p h}{\p \vr} {} & = \sint \rd \bvpp \, \Ftot (r , \bvpp) \, \frac{\p }{\p \vr} \bigg( \frac{1}{|\bv - \bvpp|} \bigg) ,
\\
\frac{\p h}{\p \vt} {} & = \sint \rd \bvpp \, \Ftot (r , \bvpp) \, \frac{\p }{\p \vt} \bigg( \frac{1}{|\bv - \bvpp|} \bigg) .
\end{align}
\end{subequations}
Here, we have $|\bv-\bvpp|^2=w_{x}^2+w_{y} ^2+w_{z}^2$
with
\begin{equation}
w_{ x}  =\vt - \vppx ,\quad
w_{ y} =  - \vppy ,\quad
w_{ z}  =\vr-\vppz ,
\label{def_w}
\end{equation}
so that the radial and tangential derivative in equation~\eqref{eq:calc_grad_h}
can be computed as
\begin{subequations}
\label{eq:calc_grad_h_II}
\begin{align}
\frac{\p h}{\p \vr} {} & = \sint \rd \bvpp \, \Ftot (r , \bvpp) \, \frac{- (\vr - \vppz)}{|\bv - \bvpp|^{3}} ,
\\
\frac{\p h}{\p \vt} {} & = \sint \rd \bvpp \, \Ftot (r , \bvpp) \, \frac{- (\vt -  \vppx)}{|\bv - \bvpp|^{3}} .
\end{align}
\end{subequations}
The last step of the calculation is
to perform the change of integration variables $\bvpp \to \bw$,
and use equations~\eqref{def_bw} and~\eqref{def_w}.
We obtain
\begin{subequations}
\label{eq:calc_grad_h_III}
\begin{align}
\frac{\p h}{\p \vr} {} & = -\sint\rd w \rd \theta \rd \phi \, \sin\theta \cos \theta \,\Ftot(r,\bv-\bw) ,
\\
\frac{\p h}{\p \vt} {} & = - \sint \rd w \rd \theta\rd \phi \, \sin^2 \theta \cos \phi \, \Ftot(r,\bv-\bw) .
\end{align}
\end{subequations}
The same method can be applied
for the gradients of $g (r , \vr, \vt)$.
Starting from equation~\eqref{eq:def_gh},
similarly to equation~\eqref{eq:calc_grad_h_II},
we obtain
\begin{subequations}
\label{eq:calc_grad_g}
\begin{align}
\frac{\partial g}{\partial \vr} {} & = \sint \rd \bvpp \Ftot(r,\bvpp) \frac{\vr \!-\! \vppz}{|\bv \!-\!\bvpp| } ,
\\
\frac{\partial g}{\partial \vt} {} & = \sint \rd \bvpp \, \Ftot(r,\bvpp) \frac{\vt \!-\! \vppx}{|\bv \!-\! \bvpp| } .
\end{align}
\end{subequations}
Following the same step as in~\eqref{eq:calc_grad_h_III},
this becomes
\begin{subequations}
\label{eq:calc_grad_g_II}
\begin{align}
\frac{\p g}{\p \vr} {} & = \sint\rd w\rd \theta\rd \phi \,w^2 \sin \theta \cos \theta \,\Ftot(r,\bv-\bw) ,
\\
\frac{\p g}{\p \vt} {} & = \sint\rd w\rd \theta\rd \phi \, w^2 \sin^2 \theta  \cos\phi  \,\Ftot(r,\bv-\bw) .
\end{align}
\end{subequations}
To compute the second-order gradients of $g$,
we differentiate equation~\eqref{eq:calc_grad_g}
once more to get
\begin{subequations}
\begin{align}
\frac{\partial^2 g}{\partial \vr^2} &= h -\sint \rd \bvpp \,\Ftot(r,\bvpp) \frac{ (\vr-\vppz)^2  }{|\bv-\bvpp|^3 } ,
\\
\frac{\partial^2 g}{\partial \vt \partial \vr} &= -\sint \rd \bvpp \,\Ftot(r,\bvpp) \frac{(\vr-\vppz)(\vt-\vppx)}{|\bv-\bvpp|^3 } ,
\\
\frac{\partial^2 g}{\partial \vt^2}&= h -\sint\rd \bvpp \,\Ftot(r,\bvpp) \frac{(\vt-\vppx)^2 }{|\bv-\bvpp| ^3} .
\end{align}
\end{subequations}
Using once again the spherical coordinates
from Fig.~\ref{fig:Coord_syst_3}, this gives
\begin{subequations}
\label{eq:calc_grad_g_III}
\begin{align}
\frac{\partial^2 g}{\partial \vr^2} {} & \!=\! h \!-\! \sint \rd w\rd \theta\rd \phi  \,w  \sin \theta \cos^2 \theta  \,\Ftot(r,\bv \!-\! \bw) ,
\\
\frac{\partial^2 \! g}{\partial \vt \partial \vr} \!\! {} & =\!\! - \!\sint\! \rd w \rd \theta \rd \phi\,w  \sin^2 \! \theta \cos \theta \cos \phi \Ftot(r,\bv\!-\!\bw) ,
\\
\frac{\partial^2 g}{\partial \vt^2} {} & \!=\! h \!-\! \sint\rd w\rd \theta \rd \phi  \,w  \sin^3 \theta \cos^2 \phi \,\Ftot(r,\bv \!-\! \bw) ,
\end{align}
\end{subequations}
where the value of $h(r,\vr,\vt)$ follows from equation~\eqref{eq:gh_frame}.

Equations~\eqref{eq:gh_frame}--\eqref{eq:calc_grad_g_III}
are the main results of this Appendix.
We emphasise that these expressions
do not involve any diverging velocity denominators,
nor any gradients of the cluster's DF, $\Ftot$.

\subsection{Symmetries}
\label{subsec:Rosenbltuh_properties}

Focusing again on equations~\eqref{eq:vrpvtp_def} and~\eqref{eq:Ea_La},
we find that $\Epp$ and $\Lpp$ are left unchanged
by the transformation ${ (\vr , \theta) \!\to\! (- \vr , \pi - \theta) }$.
As a consequence, we get from equation~\eqref{eq:gh_frame}
that $h$ and $g$ are both even functions in $\vr$.
From the same equations~\eqref{eq:vrpvtp_def} and~\eqref{eq:Ea_La},
we also get that for $\vt = 0$, $\Epp$ and $\Lpp$
are independent of $\phi$.
When used in equation~\eqref{eq:gh_frame}
to perform the obvious integral over $\phi$,
this imposes that $\p g / \p \vt$ vanishes for $\vt = 0$.

\subsection{Integration strategy}

In equations~\eqref{eq:gh_frame}--\eqref{eq:calc_grad_g_III},
the integration bounds are naturally
set by the constraints
$0 \leq w < + \infty$,
$0 \leq \theta \leq \pi$,
and $0 \leq \phi \leq 2 \pi$.
This integration domain
may be further constrained
by using the fact that the DF $\Ftot(r,\bv-\bw) = \Ftot(E',L')$ in the integrand 
is non-zero only for bound perturbing orbits.

Expanding equation~\eqref{eq:Ea_La} yields
\begin{align}
E' {} & = \psi(r) + \tfrac{1}{2}(v^2+w^2) - w (\vr \cos \theta + \vt \sin \theta \cos \phi).
\nonumber
\\
{} & = \tfrac{1}{2} w^{2} - w (\vr \cos \theta + \vt \sin \theta \cos \phi) + E ,
\label{eq:Ep_w}
\end{align}
where we introduced $E = \psi(r) + \tfrac{1}{2}v^{2}$
the test star's energy.
We now want to restrain our integration domain to $E' < 0$,
i.e.\ to bound perturbing orbits.
At fixed $(\theta,\phi)$, the r.h.s of equation~\eqref{eq:Ep_w}
is a polynomial in $w$ with positive leading coefficient.
This polynomial is therefore always positive if it has no root,
or only negative between its roots if it has some.
Its discriminant is
\begin{equation}
\Delta = (\vr \cos \theta + \vt \sin \theta \cos \phi)^2 - 2 E\,.
\end{equation}
In practice, we evaluate diffusion coefficients
only for bound test stars,
i.e.\ test stars satisfying $E < 0$,
from which we get $\Delta > 0$.
The two roots of the polynomial are given by
\begin{subequations}
\begin{align}
w_{+} &= \vr \cos \theta + \vt \sin \theta \cos \phi + \sqrt{\Delta} > 0, \\
w_{-} &= \vr \cos \theta + \vt \sin \theta \cos \phi - \sqrt{\Delta} < 0.
\end{align}
\end{subequations}
As a consequence, we fix our integration domains
to $0 \leq \theta \leq \pi$,
$0 \leq \phi \leq 2 \pi$,
and $0 \leq w \leq w_{+} (\theta , \phi)$.
In practice, the whole integration
is performed using the standard midpoint rule
within each of the respective allowed domains.
We typically use $N_K = 10^{2}$ sampling nodes
for the computations presented throughout the main text
(see also Fig.~\ref{fig:Landau}).

\section{Coefficients from Landau}
\label{sec:Chavanis4D}

The local diffusion coefficients presented
in equation~\eqref{eq:generic_D} are expressed
as a function of the Rosenbluth potentials.
The same diffusion coefficients can also be obtained
from the homogeneous Landau equation~\citep[see][for a review]{Chavanis2013}.
In this Appendix, we start from this equivalent writing
to check our calculations and numerical implementations.

Starting from equations~{(F.3)--(F.6)} in~\cite{Chavanis2013},
we consider the local diffusion coefficients
\begin{subequations}
\label{eq:ChavanisNR}
\begin{align}
{} & \hskip -0.2cm\langle \Delta v_i \rangle \!=\! \!\half (2\pi)^{4} (\!m \!+\!\mb\!) \sint\rd \bvpp \!\rd \bk\, k_i k_j  \deltaD \!(\!\bk \! \cdot \! \bw \!)\hu (k)^2 \!\frac{\partial \Ftot}{\partial \vpp_j} ,
\\
{} &\hskip -0.2cm\langle \Delta v_i  \Delta v_j \rangle \!=\! (2\pi)^{4} \mb \sint\rd \bvpp \!\rd \bk \,k_i k_j \deltaD(\bk \! \cdot \! \bw) \hu (k)^2 \Ftot ,
\end{align}
\end{subequations}
with $\bw\!=\!\bv-\bvpp$ standing
for the velocity difference between the test and background particles.
In that expression, we also introduced
$(2\pi)^3 \hu (k)\!=\!-4\pi G/k^2$
as the Fourier transform of the gravitational potential,
while the DF, $\Ftot$, and its derivatives are evaluated at $(\br,\bvpp)$.

Our goal is now to further simplify equation~\eqref{eq:ChavanisNR}
by using the same frame as in Fig.~\ref{fig:Coord_syst_1_spe}.
We first introduce spherical coordinates $(k , \theta_{k} , \phi_{k})$
to write the radial frequency, $\bk$, as
\begin{equation}
k_{1} \!= \! k\cos\theta_k ,\quad  \! \!
k_{2}  \!= \! k\sin\theta_k\cos\phi_k , \quad  \! \!
k_{3} \! = \! k\sin\theta_k\sin\phi_k .  \! \!
\label{k_coordinate}
\end{equation}
In order to make progress with equation~\eqref{eq:ChavanisNR},
we must deal with the resonance condition
$\deltaD (\bk \cdot \bw)$. With the present frame,
it imposes the cancellation of
\begin{equation}
\bk \cdot \bw = k (v-\vpp_{1}) \cos \theta_k -k\sin \theta_k(\vpp_{2}\cos \phi_k + \vpp_{3}\sin \phi_k) .
\label{eq:res_cond}
\end{equation}
We note that $\bk \cdot \bw$ is generically a monotonic function
of $\vpp_{i}$ when varied individually.
As such, if one decides to solve the resonance
condition with respect to a given $\vpp_{i}$,
one can use the relation
\begin{equation}
\deltaD (\bk \cdot \bw) = \frac{\deltaD (\vpp_{i} - \vpp_{i,\res})}{|\p (\bk \cdot \bw) / \p \vpp_{i}|_{\vpp_{i} = \vpp_{i,\res}}} ,
\label{solve_res}
\end{equation}
where $\vpp_{i,\res}$ stands for the (single) root
of the function ${ \vpp_{i,\res} \mapsto \bk \cdot \bw }$.
From equation~\eqref{eq:res_cond},
we generically find that, depending on the considered coordinate,
the resonance condition can generically be solved as
\begin{subequations}
\label{eq:sol_res}
\begin{align}
\vpp_{1,\res}&=v - \tan \theta_k(\vpp_{2}\cos \phi_k + \vpp_{3}\sin \phi_k) ,
\\
 \vpp_{2,\res}&= \frac{(v-\vpp_{1})\cot \theta_k -\vpp_{3}\sin \phi_k }{\cos \phi_k} ,
 \\
 \vpp_{3,\res}&=  \frac{(v-\vpp_{1})\cot \theta_k-\vpp_{2}\cos \phi_k }{\sin \phi_k} ,
\end{align}
\end{subequations}
with the associated gradients
\begin{subequations}
\label{eq:grad_res}
\begin{align}
\frac{\partial (\bk \cdot \bw )}{\partial \vpp_1} &= - k \cos \theta_k,\\
\frac{\partial (\bk \cdot \bw )}{\partial \vpp_2} &= - k \sin \theta_k \cos \phi_k,\\
\frac{\partial (\bk \cdot \bw )}{\partial \vpp_3} &= - k \sin \theta_k \sin \phi_k .
\end{align}
\end{subequations}
Let us now illustrate the calculation of $\langle (\Delta v_{1})^{2} \rangle$.
Starting from equation~\eqref{eq:ChavanisNR},
we write
\begin{align}
\langle (\Delta v_1)^2 \rangle {} &=(2\pi)^4 \mb \sint \rd \bvpp \rd \bk \,k_1^2 \,\deltaD(\bk \! \cdot \! \bw) \hu (k)^2 \, \Ftot
\label{eq:Deltvav1_Chavanis}
\\
{} & = 4 \mb G^2\sint \rd \bvpp \rd k  \rd \theta_k \rd \phi_k \, \sin \theta_k \cos^2\theta_k\, \deltaD(\bk \! \cdot \! \bw) \Ftot .
\nonumber
\end{align}
Following equation~\eqref{solve_res},
we use $\vpp_{1}$ to solve the resonance condition
and obtain
\begin{align}
\hskip -0.2cm \langle (\Delta v_1\!)^2 \rangle {} &  \!=\! 4 \mb  G^2 \!\sint \rd \bvpp \! \rd k  \rd \theta_k \rd \phi_k \!\frac{ \sin \theta_k \! \cos^2 \!\theta_k}{|k \cos \theta_k|} \deltaD \!(\vpp_{1} \!\!-\!\vpp_{1,\res} \!) \Ftot 
\notag 
\\
{} & \!=\! 4 \mb G^{2} \sint\rd \bvpp \! \frac{\rd k}{k} \rd \theta_{k} \rd \phi_{k} \!\sin \theta_k \!|\! \cos \theta_k \!| \deltaD \!(\vpp_{1} \!\!-\! \vpp_{1 , \res}\!) \Ftot .
\label{eq:calc_Deltav1_Chavanis}
\end{align}
Importantly, we emphasise that using here $\vpp_{1}$
to solve the resonance condition naturally leads
to the simplification of any diverging denominators.
As expected, we must heuristically cure the integral over radial frequency
on both small and large scales~\citep[see, e.g.\@, \S{2.6} in][]{Chavanis2013}.
Introducing the Coulomb logarithm
\begin{equation}
\ln \Lambda = \!\int_{\kmin}^{\kmax} \!\!\!\! \rd k / k = \ln \big( \kmax / \kmin \big) ,
\label{eq:def_lnLambda}
\end{equation}
equation~\eqref{eq:calc_Deltav1_Chavanis} finally becomes
\begin{align}
 \langle (\Delta v_{1})^{2} \rangle = A \, \mb \!\int_{\vpp_1 = \vpp_{1,\res}} \hspace{-10mm} \rd \vpp_2 \rd \vpp_3 \rd \theta_k \rd \phi_k \, \sin \theta_k |\!\cos\theta_k \!| \, \Ftot ,
\end{align}
with $A = 4 G^2  \ln \Lambda$
and $\Ftot = \Ftot (\br , \bvpp)$.
In that expression, the subscript ${ \vpp_{1} = \vpp_{1,\res} }$
attached to the integral symbol implies
that the resonance condition from equation~\eqref{eq:sol_res}
has been applied to the variable $\vpp_{1}$.

Following equation~\eqref{eq:SpecialCoordSys_LocDiffCoeffs},
we can use the exact same approach to compute
the parallel and perpendicular diffusion coefficients.
During that calculation,
we pay attention to solving the resonance condition
$\deltaD (\bk \cdot \bw)$
with respect to the appropriate velocity, $\vpp_{i}$,
to ensure that the diverging denominators naturally cancel out.
Overall, one gets
\begin{subequations}
\label{eq:4D_dvParSqAvrLoc}
\begin{align}
& \dvParAvrLoc \!=\! A \half (m \!+\! \mb)
\\
& \hskip 0.8cm\times \bigg[\!\int_{\vpp_1 = \vpp_{1,\res}} \hspace{-11mm} \rd \vpp_2 \rd \vpp_3 \rd \theta_k \rd \phi_k \, |\!\cos \theta_k \!|  \sin \theta_k \frac{\partial \Ftot}{\partial \vpp_1}
\notag
\\
& \hskip 1.0cm+  \!\int_{\vpp_2 = \vpp_{2,\res}} \hspace{-11mm} \rd \vpp_1 \rd \vpp_3 \rd \theta_k \rd \phi_k \cos \theta_k \sin \theta_k \, \sgn(\cos \phi_k) \frac{\partial \Ftot }{\partial \vpp_2}
\notag
\\
& \hskip 1.0cm+ \!\int_{\vpp_3 = \vpp_{3,\res}}\hspace{-11mm} \rd \vpp_1 \rd \vpp_2 \rd \theta_k \rd \phi_k\,  \cos \theta_k  \sin \theta_k \, \sgn(\sin \phi_k) \frac{\partial \Ftot }{\partial \vpp_3} \bigg] ,
\notag
\\
&  \dvParSqAvrLoc \!=\!  A \mb \!\!\int_{\vpp_1 = \vpp_{1,\res}} \hspace{-11mm} \rd \vpp_2 \rd \vpp_3 \rd \theta_k \rd \phi_k |\! \cos \theta_k \!| \sin \theta_k \Ftot  , 
 \\
 &   \dvPerpSqAvrLoc \!=\! A \mb \!\bigg[\!\int_{\vpp_2 = \vpp_{2,\res}} \hspace{-11mm} \rd \vpp_1 \rd \vpp_3  \rd \theta_k \rd \phi_k \sin^{2} \!\theta_k |\! \cos \phi_k \!| \Ftot 
  \\
  &  \hskip 2.1cm+  \!\int_{\vpp_3 = \vpp_{3,\res}}  \hspace{-11mm}  \rd \vpp_1 \rd \vpp_2 \rd \theta_k \rd \phi_k \sin^{2} \theta_k |\! \sin \phi_k \!| \Ftot \bigg] ,
  \nonumber
  \end{align}
\end{subequations}
with $\sgn(x)=x/|x|$.

The integration domains in equation~\eqref{eq:4D_dvParSqAvrLoc} are respectively given by
$0 \leq \theta_k \leq \pi$ and $0 \leq \phi_k \leq 2\pi$ for the angular variables, and $-\sqrt{-2\psi(r)} \leq \vpp_i \leq \sqrt{-2\psi(r)}$ for the velocity ones.
In equation~\eqref{eq:4D_dvParSqAvrLoc},
the DF and its derivatives are evaluated in $(r,\bvpp)$, or equivalently in  $(E',L')$ given by
\begin{subequations}
\begin{align}
& \Epp \!=\! \psi(r)+\half \vppSq ,
  \;
  \Lpp \!=\! r \vppt  ,
  \\
  & \vppr \!=\! \frac{\vpp_1 \vr}{v} \!+\! \frac{\vpp_2 \vt}{v} ,
  \;
  \vppt \!=\! \sqrt{\vppSq \!-\! \vpprSq}  ,
  \;
  {\vppSq} \!\!=\! \vppSq_1 \!\!+\! \vppSq_2 \!\!+\! \vppSq_3 .
\end{align}
\end{subequations}
Furthermore, all the derivatives of $\Ftot = \Ftot (\Epp,\Lpp)$
with respect to $\vpp_{i}$
are computed using the chain rule so that
\begin{equation}
\frac{\p \Ftot}{\p \vpp_i} = \frac{\p E'}{\p \vpp_i} \frac{\p \Ftot}{\p E'} + \frac{\p L'}{\p \vpp_i} \frac{\p \Ftot}{\p L'},
\end{equation}
where, following Fig.~\ref{fig:Coord_syst_1_spe}
and equation~\eqref{eq:Ea_La},
the remaining gradients simply read
\begin{subequations}
\begin{align}
\frac{\p E'}{\p \vpp_i} &= \vpp_i,\quad \frac{\p L'}{\p \vpp_i}  =r \bigg(\frac{\vpp_i}{\vppt} - \frac{\vppr}{\vppt} \frac{\p \vppr}{\p\vpp_i}\bigg),\\
\frac{\p \vppr}{\p\vpp_i}  &= \begin{cases}
\displaystyle{\vr/v}, & {\mathrm{if}} \ i=1 ,\\
\displaystyle{\vt/v},&{\mathrm{if}}\ i=2 , \\
\displaystyle{0},&{\mathrm{if}}\ i=3 .
\end{cases}
\end{align}
\end{subequations}
Equation~\eqref{eq:4D_dvParSqAvrLoc}
is the main result of this Appendix.
It captures the exact same velocity diffusion coefficients as
equation~(\ref{eq:generic_grad}) from the main text.
Yet, these expressions
require the computation of four-dimensional integrals,
and involve derivatives of the cluster's DF.
These two difficulties are lifted by 
using Rosenbluth's approach as the final expressions of the potentials only 
involve three-dimensional integrals
without any occurrences of derivatives
of the cluster's DF,
see equations~\eqref{eq:gh_frame}--\eqref{eq:calc_grad_g_III}.

\begin{figure}
\centering
\includegraphics[width=0.45 \textwidth]{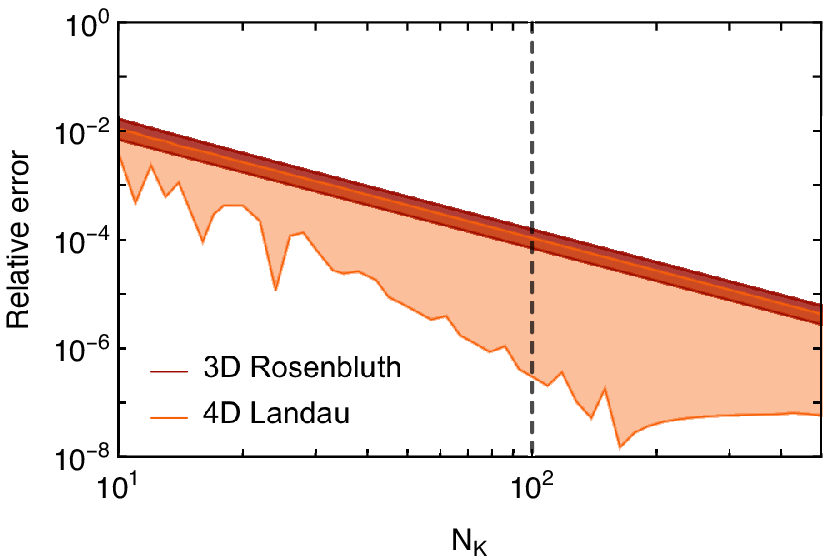}
\caption{Relative errors in $\dvParSqAvrLoc (r,\vr,\vt)$
as computed via the Rosenbluth expressions
(equation~\ref{eq:generic_grad})
or the Landau expressions (equation~\ref{eq:4D_dvParSqAvrLoc}).
Both expressions are computed using a midpoint sampling
with $N_K$ nodes.
Relative errors are computed with respect
to the Rosenbluth expressions using $N_K=5\,000$.
The level lines correspond to the 16\% and 84\% centiles
over 160 values of $(r,\vr,\vt)$
drawn uniformly
within $0 \leq r \leq b$,
$0 \leq \vr \leq \sqrt{-2\psi (r)}$ and $0 \leq \vt \leq \sqrt{-2\psi (r)-\vr^{2}}$.
Both approaches are found to be in agreement.
}
\label{fig:Landau}
\end{figure}

In Fig.~\ref{fig:Landau}, we illustrate
the relative error between the Rosenbluth expressions
(equation~\ref{eq:generic_grad})
and the Landau ones (equation~\ref{eq:4D_dvParSqAvrLoc}),
as a function of the number of nodes, $N_{K}$,
used in the midpoint rule. 
Reassuringly, we find that
(i) both expressions are in agreement;
(ii) the use of the midpoint rule ensures
a convergence in $O (N_{K}^{-2})$.
In practice, we used
$N_K=10^{2}$ sampling nodes
in the main text,
ensuring a relative error of the order of $10^{-4}$.

\section{Distribution function}
\label{sec:DF}

\begin{figure*}
\centering
\includegraphics[width=1.0 \textwidth]{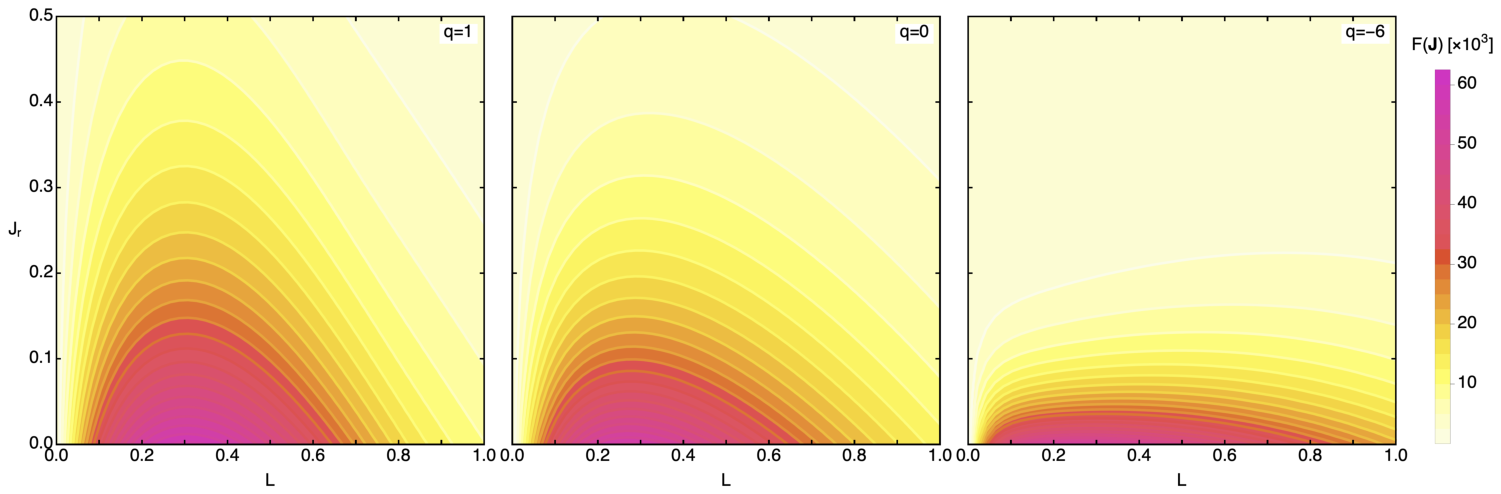}
\caption{Illustration of the reduced DF
from equation~\eqref{eq:def_Fq}
for a radially anisotropic distribution ($q=1$, left),
an isotropic one ($q=0$, middle),
and a tangentially anisotropic one ($q=-6$, right).
The radially (resp.\ tangentially) anisotropic DF
shows a higher concentration
of radial (resp.\ tangential) orbits,
i.e.\ orbits with small $L$ (resp.\ small $\Jr$).
}
\label{fig:DF}
\end{figure*}

To investigate the effects associated with velocity anisotropies,
we consider the families of DFs put forward by~\cite{Dejonghe1987}
\begin{equation}
\Ftot (E,L;q) = \frac{M}{\Lo^{3}}\frac{3 \, \Gamma (6 \!-\! q) \, \tE^{7/2-q}}{2(2\pi)^{5/2}} \, \mH_q \bigg(\frac{\tL^{2}}{2\tE}\bigg) ,
\label{eq:def_Fq}
\end{equation}
where $q$ controls the flavour and degree of anisotropy
in the cluster (see Fig.~\ref{fig:beta}).
In equation~\eqref{eq:def_Fq},
we introduced the rescaled energy and angular momentum,
$\tE = E/\Eo$ and $\tL=L/\Lo$,
with
\begin{equation}
\Eo=-GM/b ,
\quad
\Lo=\sqrt{GMb}.
\label{eq:defE0L0}
\end{equation}
We also introduced the function
\begin{equation}
\mathbb{H}_q(x) \!=\! \begin{cases}
\displaystyle{\frac{\!\HG(\half q,q \!-\!\tfrac{7}{2},1;x)}{\Gamma(\tfrac{9}{2} \!-\! q)} }, & {\mathrm{if}} \ x\leq1 ,
\\[2.0ex]
\displaystyle{\frac{\!\HG(\half q, \half q, \half (9 \!-\! q);1/x)}{\Gamma(1 \!-\! \half q)\Gamma(\half (9 \!-\! q))} } \frac{1}{x^{q/2}},&{\mathrm{if}}\ x \geq1 ,
\end{cases}
\end{equation}
with $\HG$ the hypergeometric function
and $\Gamma$ the Gamma function.
In Fig.~\ref{fig:DF},
we illustrate the reduced DF, $F = 2 L \Ftot$
(see equation~\ref{def_Fred}),
in action space for various anisotropies.

\section{Orbit-averaging}
\label{sec:orbit-average}
In order to orbit-average the local diffusion coefficients, we 
proceed in three steps:
(i) we compute the local diffusion coefficients
in the orbit's invariants $E$ and $L$;
(ii) we carry out the orbit average using an explicit
effective anomaly which regularises the process;
(iii) we compute the
corresponding orbit-averaged diffusion coefficients in terms
of the actions, $\Jr$ and $L$.

\subsection{From velocity to energy and angular momentum}
\label{subsec:vel_to_energy}

Equation~\eqref{eq:generic_grad} gives
the local velocity diffusion coefficients.
Following equation~{(C15)}--{(C19)} of~\citet{BarOr2016},
these can be translated into local diffusion coefficients
in energy and angular momentum via
\begin{subequations}
\label{eq:DE_DL}
\begin{align}
\DEAvr & = \half \dvParSqAvrLoc + \half \dvPerpSqAvrLoc+v\dvParAvrLoc ,
\\
\DESqAvr & =v^{2}\dvParSqAvrLoc ,
\\
\DLAvr& =\frac{L}{v}\dvParAvrLoc + \tfrac{1}{4} \frac{r^{2}}{L}\dvPerpSqAvrLoc ,
\\
\DLSqAvr & = \frac{L^{2}}{v^{2}}\dvParSqAvrLoc + \half \big(r^{2} \!-\!(L/v)^{2} \big) \dvPerpSqAvrLoc ,
\\
\DEDLAvr & = L \dvParSqAvrLoc .
\end{align}
\end{subequations}
To evaluate equation~\eqref{eq:DE_DL},
we must switch between $(r,\vr,\vt)$ and $(E,L)$.
This is naturally done through
\begin{equation}
E = \psi(r) + \half \vr^2 + \half \vt^2 ,
\quad
L = r \, \vt ,
\end{equation}
and the associated inverse transformation
\begin{equation}
\vr = \pm \sqrt{2(E-\psi(r))-L^2/r^2 } ,
\quad
\vt = L/r .
\label{eq:E_to_vr}
\end{equation}
In equation~\eqref{eq:E_to_vr},
the ``$+$'' case for $\vr$ corresponds
to the outward radial motion from pericentre to apocentre,
while the ``$-$'' case corresponds to the inward motion.
In practice, as the Rosenbluth potentials
$h$ and $g$ are even in $\vr$
(see~\S\ref{subsec:Rosenbltuh_properties}),
we can safely limit ourselves to the outward part of the orbit,
i.e.\ $\vr > 0$, when orbit-averaging.

\subsection{Plummer effective anomaly}
\label{subsec:PlummerEffAnomaly}

Once the local diffusion coefficients in $(E,L)$ estimated,
they must be averaged along the test star's orbit,
as in equation~\eqref{eq:orbit_avg}.
Following~\cite{Henon1971},
we perform the orbit-average with respect
to an effective anomaly $-1 \leq u \leq 1 $,
such that $r (u = - 1) = \rrp $
and $r (u = 1) = \rra$.
Limiting ourselves to the outward
part of the orbit,
we rewrite equation~\eqref{eq:orbit_avg} as
\begin{equation}
D_{E} = \frac{2}{ T} \! \int_{-1}^{1} \!\! \rd u \, \Theta (u) \, \langle \Delta E \rangle ,
\label{rewrite_avg}
\end{equation}
with the  weight function
\begin{equation}
\Theta (u) =  \frac{1}{\vr} \frac{\rd r}{\rd u} .
\label{def_Theta}
\end{equation}
Taking inspiration from~\S{G} of~\cite{Fouvry+2021},
for the Plummer potential,
we take our effective anomaly, $u \mapsto r (u)$, to be
\begin{equation}
r (u) = b \, \sqrt{s^{2} (u) - 1} ,
\label{def_r_eff}
\end{equation}
where
\begin{equation}
s (u) = a \big( 1 + e f (u) \big) ,\quad
\label{def_s}
\mathrm{ with }\quad
f(u) = u \, \big( \tfrac{3}{2} - \half u^{2} \big).
\end{equation}
In that expression, the effective semi-major axis, $a$,
and eccentricity, $e$, are defined as
\begin{equation}
a = \frac{\sra + \srp}{2} ,
\quad
e = \frac{\sra - \srp}{\sra + \srp} ,
\label{def_a_e_eff}
\end{equation}
with $\srp = \sqrt{1 + (\rrp/b)^{2} }$
and similarly for $\sra$.
For a given orbit $(E,L)$, the bounds $(\srp,\sra)$
can be easily computed, as shown in equation~\eqref{eq:poly3_sp_sa}.
The key motivation of the  choice of equation~\eqref{def_r_eff},
is that the weight function, $\Theta (u)$  given by equation~\eqref{def_Theta}
becomes explicit and is numerically well-behaved
for all $-1 \leq u \leq 1$.
Indeed, after simplification, one gets
\begin{equation}
\Theta (u) = \frac{1}{\Omegao}\frac{3}{4 \sqrt{2} }\frac{\sqrt{\sra \srp (\sra + \srp) }}{\sqrt{4 -
u^2}} \frac{  \mathcal{A}(u)^{3/2}}{
 \sqrt{ \sra \srp \mathcal{A}(u) +\mathcal{B}(u) }} ,
\label{final_Theta}
\end{equation}
with the frequency scale $\Omegao = \sqrt{G M / b^{3}}$.
In equation~\eqref{final_Theta},
we also introduced
\begin{align*}
\mathcal{A}(u) &=\srp (u+2) (u-1)^2- \sra ( u-2) (u+1)^2 ,
\\
\mathcal{B}(u) &= \srp ( u^3- 3 u+6 )-\sra ( u^3- 3 u-6) ,
\end{align*}
which are both always positive.
The mapping from equation~\eqref{def_r_eff}
is also used to compute the radial period $T$
via
\begin{equation}
\frac{T}{2} = \! \int_{-1}^{1} \!\! \rd u \,  \Theta (u).
\label{calc_T}
\end{equation}
In practice, to compute the orbit average
from equation~\eqref{rewrite_avg}
we use a midpoint rule with $\Navg = 10^{2}$ nodes,
for which the typical relative error is found to be $10^{-5}$.

\subsection{Action space diffusion coefficients}
\label{subsec:change_var}

The orbit-averaged diffusion coefficients
in $(E,L)$-space can finally be converted
into action space $\bJ = (\Jr , L)$.
To do so, we follow the generic change of coordinates
presented in equations~{(122)} and~{(123)} of~\cite{BarOr2016},
and write
\begin{subequations}
\label{eq:orb_avg_change_var}
\begin{align}
  D_{\Jr} & \!=\! \frac{\partial \Jr}{\partial E }D_{E}+\frac{\partial \Jr}{\partial L}D_{L}
  \nonumber
  \\ 
  & + \half \frac{\partial^{2}\Jr}{\partial E^2}D_{EE} + \half \frac{\partial^{2}\Jr}{\partial L^2}D_{LL}+\frac{\partial^{2}\Jr}{\partial E\partial L} D_{EL},
  \\
D_{\Jr L} & \!=\! \frac{\partial \Jr}{\partial E}D_{EL} +\frac{\partial \Jr}{\partial L}D_{LL},
\\
D_{\Jr \Jr} & \!=\! \bigg(\!\frac{\partial \Jr}{\partial E} \!\bigg)^2 \!\! D_{EE} \!+\! 2 \, \frac{\partial \Jr}{\partial E} \frac{\partial \Jr}{\partial L}D_{EL} \!+\! \bigg(\! \frac{\partial \Jr}{\partial L} \!\bigg)^2\!\! D_{LL} . 
\end{align}
\end{subequations}
Equation~\eqref{eq:orb_avg_change_var}
requires the gradients of $\Jr$
with respect to $(E,L)$.
To proceed, we start from the integral definition
of $\Jr$~\citep[see, e.g.\@, equation~{3.224} in][]{Binney2008}
\begin{equation}
\Jr = \frac{1}{\pi} \!\int_{\rrp}^{\rra} \rd r \, \vr= \frac{1}{\pi } \!\int_{-1}^{1}\rd u\, \Theta(u) \, \vr^2,
\label{eq:Jr_def}
\end{equation}
with the radial velocity $\vr^{2} = 2(E-\psi(r))-L^2/r^2$,
and $\Theta (u)$ following from equation~\eqref{final_Theta}.
The first-order derivatives of equation~\eqref{eq:Jr_def}
are naturally given by
\begin{subequations}
\begin{align}
  \frac{\partial \Jr}{\partial E} &= \frac{1}{\pi} \!\int_{\rrp}^{\rra}\frac{\rd r}{\vr} = \frac{1}{\pi} \!\int_{-1}^{1}\rd u\, \Theta(u)  = \frac{T}{2\pi} ,
  \\
  \frac{\partial \Jr}{\partial L} &=- \frac{L}{\pi} \!\int_{\rrp}^{\rra}\frac{ \rd r}{r^2 \vr}
=- \frac{L}{\pi}  \!\int_{-1}^{1} \rd u \, \frac{\Theta(u) }{r^2(u) }. 
  \end{align}
  \end{subequations}
Similarly, the  second-order derivatives read
\begin{subequations}
\label{eq:def_Jr_u}
\begin{align}
  \frac{\partial^2 \Jr}{\partial E^2}  &=  \frac{1}{\pi} \!\int_{-1}^{1} \rd u \,\frac{\partial \Theta}{\partial E} ,
  \\
  \frac{\partial^2 \Jr}{\partial L \partial E} &=   \frac{1}{\pi} \!\int_{-1}^{1} \rd u\,\frac{\partial \Theta}{\partial L} ,
  \\
  \frac{\partial^2 \Jr}{\partial L^2} &=-  \frac{1}{\pi } \bigg[\!\int_{-1}^{1}\frac{ \rd u\, \Theta(u) }{r^2(u)}
  \notag
  \\ 
  & \hskip -0.5cm + L\!\int_{-1}^{1} \frac{\rd u}{r^4(u)}\bigg( \frac{\partial \Theta}{\partial L} r^2(u) - 2 b^2 \Theta(u) s(u) \frac{\partial s}{\partial L} \bigg)  \bigg] ,
  \end{align}
\end{subequations}
with $s(u)$ given by equation~\eqref{def_s}.

All the gradients appearing in equation~\eqref{eq:def_Jr_u}
can be obtained using the chain rule
and implicit differentiation.
Indeed, we write
\begin{equation}
  \frac{\partial \Theta}{\partial E} =  \frac{\partial \srp}{\partial E}  \frac{\partial \Theta}{\partial \srp}  + \frac{\partial \sra}{\partial E}  \frac{\partial \Theta}{\partial \sra},
  \label{eq:thetaE}
\end{equation}
and similarly for $\p \Theta / \p L$ and $\p s / \p L$.
In that expression, the gradients $\p \Theta / \p \srp$
and $\p \Theta / \p \sra$ follow from equation~\eqref{final_Theta}.
Using equations~\eqref{def_s} and~\eqref{def_a_e_eff},
we also have
\begin{equation}
  \frac{\partial s}{\partial \srp} = \half (1 \!-\! f (u)),
  \quad
    \frac{\partial s}{\partial \sra} = \half (1 \!+\! f(u)) .
\label{eq:grad_s}
\end{equation}
Now, we must compute the gradients
of $\srp$ and $\sra$ with respect to $E$ (and $L$) entering equation~\eqref{eq:thetaE}.
Energy conservation,
$\vr^2 = 2 (E - \psi (r)) - L^{2} / r^2$,
in conjunction with the mapping from equation~\eqref{def_r_eff}
imposes the constraint 
\begin{equation}
E = \frac{\Eo}{\sst} + \frac{L^2}{2b^2(\sst^2 \!-\!1)} \equiv \Phieff(\sst,L),
\label{eq:def_sp_sa}
\end{equation}
 with $\sst=\srp$ or $\sra$.
When differentiated, equation~\eqref{eq:def_sp_sa} gives
\begin{equation}
   \frac{\p \sst}{\partial E} = \frac{1}{\p \Phieff / \p \sst} ,
   \quad
    \frac{\p \sst}{\partial L} = - \frac{\p \Phieff/\p L}{\p \Phieff / \p \sst},
\end{equation}
with
  \begin{equation}
\frac{\p \Phieff}{\p \sst} =-\frac{\Eo}{\sst^2}-\frac{s_{\star} L^2  }{b^2 (\sst^2 \!-\! 1)^2} ,
\quad
\frac{\p \Phieff}{\p L} = \frac{ L}{b^2 (\sst^2 \!-\! 1)}.
\end{equation}

Finally, for a given value of $(E,L)$, we must still identify
the associated $(\srp,\sra)$.
To do so, we note that equation~\eqref{eq:def_sp_sa}
can be rewritten as a third degree polynomial in $\sst$
\begin{equation}
 \tE \, \sst^3 - \sst^2 + \big( \half \tL^2 \!-\! \tE \big) \, \sst + 1 = 0 ,
\label{eq:poly3_sp_sa}
\end{equation}
whose  three roots are  all real and involve the 
rescaled energy and angular momentum
introduced in equation~\eqref{eq:defE0L0}.
The product of these roots is $-1/\tE < 0$,
while their sum is $1/\tE > 0$.
As a consequence, this polynomial has one negative root,
and two positive ones, namely $\srp$ and $\sra$.
In practice, the search for $\srp$ and $\sra$
was performed using \href{https://github.com/giordano/PolynomialRoots.jl}{$\texttt{PolynomialRoots.jl}$}~\citep{Skowron+2012}.

\section{$N$-body simulations}
\label{sec:NBODY}

The simulations presented throughout the main text
were performed using the direct $N$-body code
\texttt{NBODY6++GPU}~\citep{Wang+2015},
version 4.1.
The initial conditions for the anisotropic Plummer spheres
(see equation~\ref{eq:def_Fq})
were generated from \texttt{PlummerPlus.py}~\citep{Breen+2017},
while we used the same input file as in~\S{H1} of~\cite{Fouvry+2021}.
Internally, \texttt{NBODY6++GPU} uses H\'enon units ($\rHU$)~\citep{Henon1971},
defined such that $G = M = \Rv = 1 \, \rHU$,
with $G$ the gravitational constant,
$M$ the cluster's total mass and $\Rv$ its virial radius.
For the Plummer potential from equation~\eqref{Plummer_potential},
one readily finds $\Rv/b = 16/(3 \pi)$~\citep[see, e.g.\@, Table~{1} p.\ {81} in][]{HeggieHut2003}.
Each $N$-body realisation was composed of $N=10^{5}$ stars
and integrated up to $\tmax = 1\,000 \, \rHU$
with a dump every $\Delta t = 1 \, \rHU$.
On a 40-core node with a single V100 GPU,
one simulation typically required
$\!\sim\! 24 \, \mathrm{h}$ of computation.
In practice, we considered anisotropies set by
$q = 1, 0, -6, -16, -30$
and, depending on the values of $q$,
performed either $\Nrun = 50, 100$
independent realisations,
as spelled out in Table~\ref{table:para_NBODY}.
\begin{table}
\centering
\setlength{\tabcolsep}{.5em}
 \begin{tabular*}{0.89\columnwidth}{@{}lccccc}
\hline
\hline
$q$ & 1 & 0  & -6 & -16 & -30 \\
\hline
$\Nrun$ & 100 & 100 & 100 & 50 & 50 \\
$\tlast [\rHU]$& 1000 & 1000 & 100 & 100 & 100  \\
$(N_{\Jr} , N_{L}) $ & (20,15) & (20,20) & (20,20) & (70,70) & (70,70) \\
$(\Jr^{\min} , \Jr^{\max})$ & (0, 0.55) & (0, 0.55) & (0, 0.6) & (0, 0.6) & (0, 0.6)  \\
$(L^{\min} , L^{\max}) $& (0,1.05) & (0,1.05) & (0,1.1) & (0,1.1) & (0,1.1) \\
\hline
\end{tabular*}
\caption{Detailed parameters for the $N$-body simulations
and the associated binnings of action space.
Following equation~\eqref{eq:estimation_F},
we binned the $\bJ = (\Jr,L)$ action space
in $N_{\Jr} \times N_{L}$ uniform bins
within the domain $\Jr^{\min} \leq \Jr \leq \Jr^{\max}$
(similarly for $L$).
All quantities are in physical units $G\!=\!M\!=\!b\!=\!1$ if not stated otherwise.
}
\label{table:para_NBODY}
\end{table}

In order to estimate the cluster's instantaneous
core radius, $\Rc (t)$, we follow the approach from~\cite{Casertano+1985}.
We use $j = 6$ neighbors to estimate the local densities
from which we compute the location
of the cluster's density centre, $\brc (t)$,
and subsequently the associated core radius, $\Rc (t)$
\citep[see equations~{(II.2)}--{(II.4)} in][]{Casertano+1985}.
In Fig.~\ref{fig:Rc}, we illustrate the evolution of $\Rc (t)$
averaged over the available realisations.

\begin{figure}
\centering
  \includegraphics[width=0.45 \textwidth]{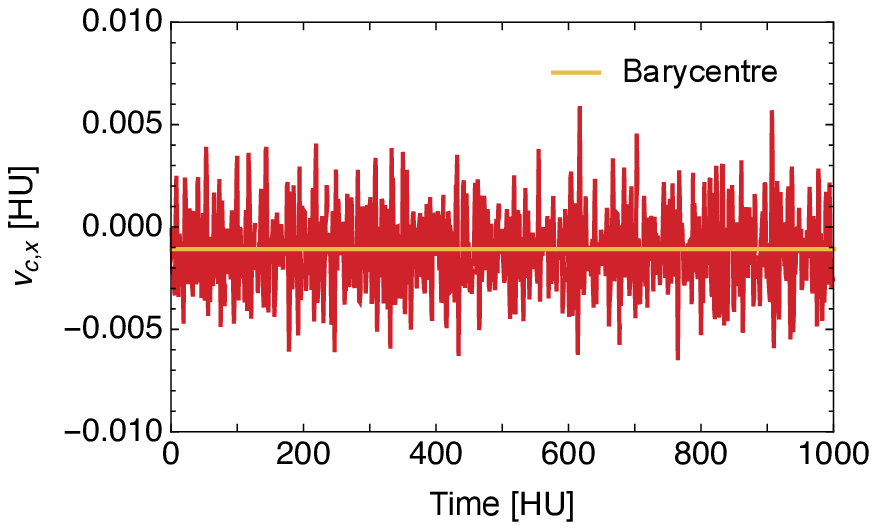}
\caption{
Time dependence of the density centre's
velocity along the $x$-direction, $v_{\rc,x}$,
for one realisation of the isotropic ($q=0$) Plummer sphere.
The density centre's velocity is found to closely follow
the barycentre uniform motion (orange line).
} 
\label{fig:density_center}
\end{figure}

In order to compute the stars' actions,
we must pick an appropriate frame.
In practice, we centre this frame
around the cluster's instantaneous density centre,
$\brc (t)$, and recentre the stars' velocities
with respect to the barycentre uniform motion.
In Fig.~\ref{fig:density_center}, we check
that indeed the density centre's instantaneous velocity,
estimated via
$ \bv_{\rc} (t) \simeq (\br_{\rc} (t \!+\! \Delta t) \!-\! \br_{\rc} (t \!-\! \Delta t))/(2\Delta t) $,
closely follows the barycentre's uniform velocity.

For a star with (recentred) position and velocity $(\br,\bv)$,
we compute its specific energy and angular momentum via
\begin{equation}
  E  = \psi(\br) + \frac{\bv^2}{2},\quad
  L = |\br \times \bv| .
\end{equation}
We subsequently keep only the bound particles,
i.e.\ particles with $E<0$.
Following equation~\eqref{eq:meanL},
the mean angular momentum, $\langle L \rangle$,
is obtained from
\begin{equation}
\langle L \rangle = \frac{1}{\Nbound}\sum_{i} L_{i} ,
\label{eq:meanL_NBODY}
\end{equation}
where $i$ runs
over all the $\Nbound$ bound particles.
Equation~\eqref{eq:meanL_NBODY}
is subsequently averaged over the available realisations
to give Fig.~\ref{fig:meanL}.

In order to compute the radial action, $\Jr$,
at a time $\tlast$,
we assume that the cluster's mean potential
does not change much from the initial Plummer profile
at $t = 0$, and integrate equation~\eqref{eq:Jr_def}
using a midpoint sampling of the effective anomaly $u$.
In order to estimate the relaxation rates,
$\p F / \p t$, presented in Fig.~\ref{fig:dFdt_NR_NBODY},
we bin the $(\Jr,L)$ action space with uniform bins.
More precisely, for a given action bin of size $\delta \Jr \!\times\! \delta L$
centered around the action $\bJ = (\Jr , L)$,
we compute
\begin{equation}
\frac{\partial F}{\partial t}(\bJ,t \!=\! 0) \simeq \frac{F(\bJ,\tlast) - F(\bJ, t \!=\! 0)}{\tlast} ,
\label{eq:estimation_F}
\end{equation}
where the local DF, $F(\bJ , t)$, follows from
\begin{align}
F(\bJ,t) &= \frac{M\, n(\bJ,t)}{(2\pi)^3 \delta \Jr \delta L},
\\
 n(\bJ,t) &= \frac{{\mathrm{stars \ in}} \ [\Jr\!-\!\tfrac{1}{2}\delta \Jr,\Jr\!+\!\tfrac{1}{2}\delta \Jr]\!\times\![L\!-\!\tfrac{1}{2}\delta L,L\!+\!\tfrac{1}{2}\delta L]}{\mathrm{total \ number \ of \ bound \ stars}}. \notag
\end{align}
Finally, we average equation~\eqref{eq:estimation_F}
over the available realisations.
As highlighted in Fig.~\ref{fig:Rc},
the stronger the tangential anisotropy,
the faster the cluster's relaxation,
and therefore the smaller the considered time $\tlast$
to ensure a minimal evolution of the cluster's mean profile.
Similarly, as illustrated in Fig.~\ref{fig:DF},
as the tangential anisotropy increases,
the DF gets more concentrated along the $\Jr = 0$ axis,
which entices us to use smaller action bins.
We detail all our binning parameters in Table~\ref{table:para_NBODY}.

In order to compute the ratios defined
in equations~\eqref{eq:Ratio_NR_NBODY} and~\eqref{eq:Ratio_NR_PIso},
we use a simple midpoint rule.
For the NR and P-Iso integrals,
given that $F$ and $\p F / \p t$ quickly drop away from the cluster's core,
we perform the integrals over the domain $0 < L < \Lo$
and $0 < \Jr < 0.5\Lo$, with $\Lo = \sqrt{G M b}$, using $200$ nodes sampled linearly in both directions.
For the $N$-body integrals,
we use the exact same bins as the ones used to estimate $\p F / \p t$
in equation~\eqref{eq:estimation_F}.

\section{More anisotropic settings}
\label{sec:more_aniso}

\begin{figure*}
\centering
\includegraphics[width=0.7\textwidth,trim = 0.2cm 0.2cm 0.1cm 0.2cm]{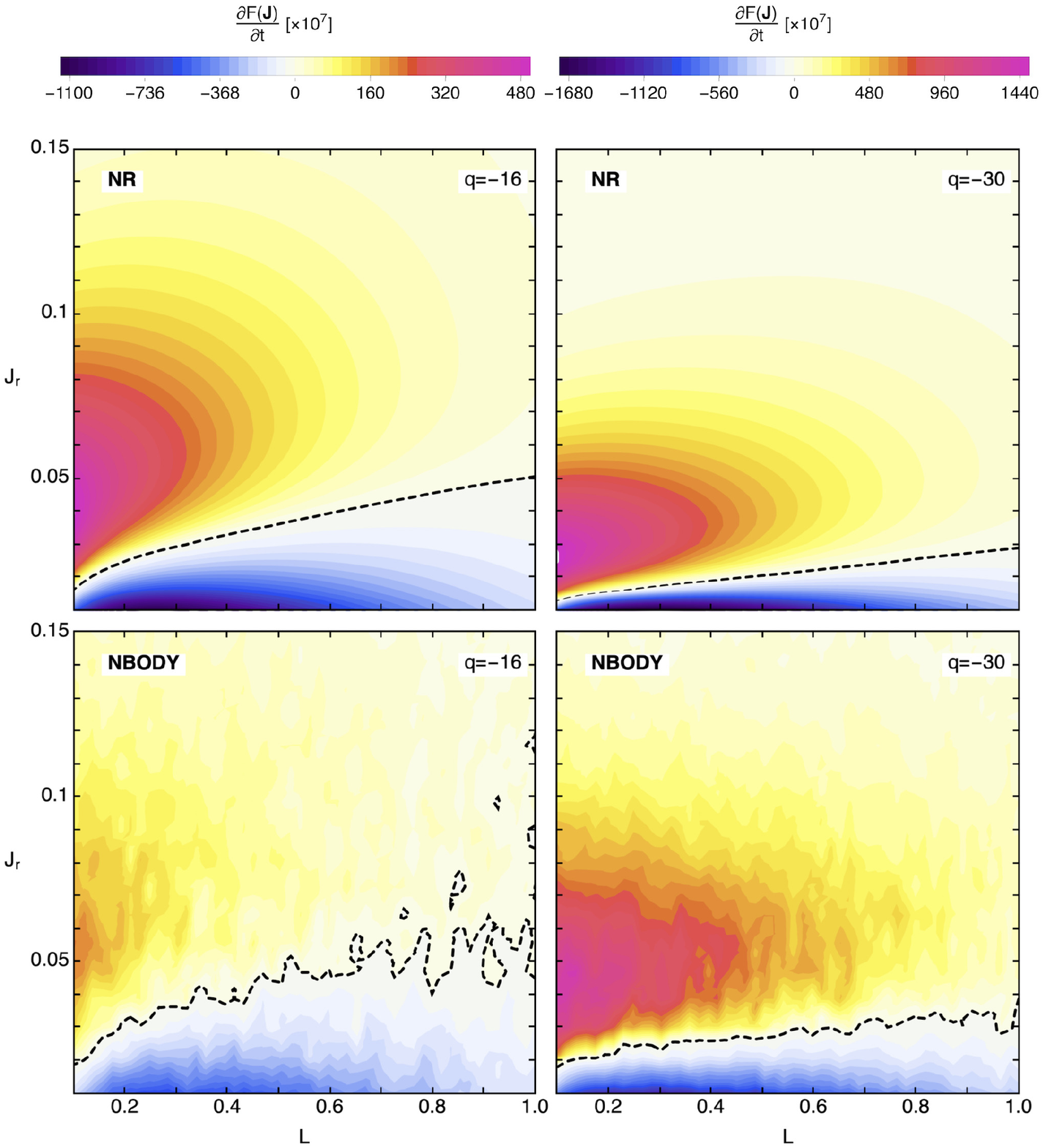}
\caption{Same as Fig.~\ref{fig:dFdt_NR_NBODY}
but for stronger tangential anisotropies.
There is a qualitative agreement
between the NR predictions and the $N$-body measurements,
up to an overall prefactor that gets larger
as the anisotropy increases (see Fig.~\ref{fig:Ratio_NR_NBODY}).
}
\label{fig:dFdt_NR_NBODY_appendix}
\end{figure*}

In this Appendix, we complement Fig.~\ref{fig:dFdt_NR_NBODY}
by considering Plummer spheres
with even stronger tangential anisotropies.
This is illustrated in Fig.~\ref{fig:dFdt_NR_NBODY_appendix}
for $q=-16,-30$.
Even in these strongly anisotropic regimes,
the anisotropic NR diffusion coefficients
from equation~\eqref{eq:generic_grad}
predict the structure of $\p F / \p t$ in action space,
up to an overall prefactor that grows
as the anisotropy increases (see Fig.~\ref{fig:Ratio_NR_NBODY}).

\section{Local contributions to diffusion}
\label{sec:pseudo_aniso}

\begin{figure}
\centering
\includegraphics[width=0.45 \textwidth]{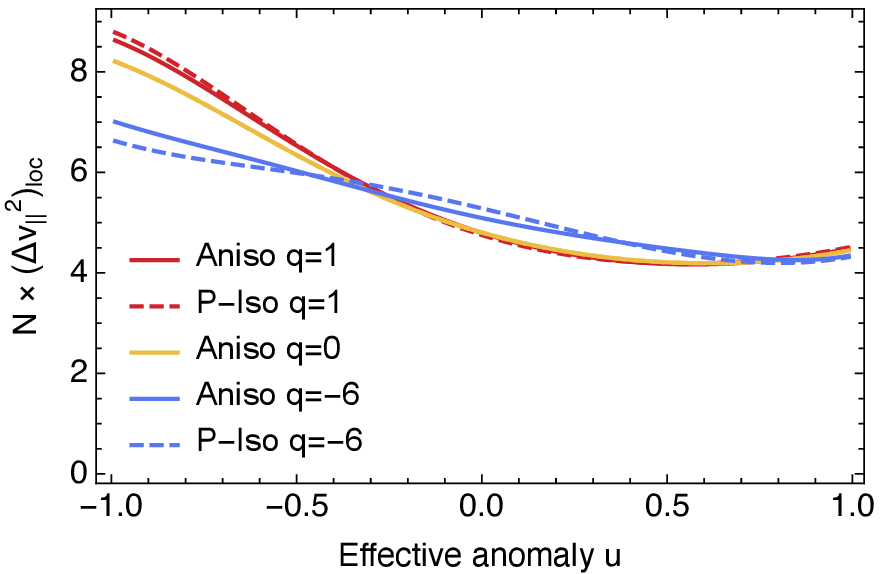}
\caption{Illustration of the velocity deflections
accumulated along the motion of a test star,
seen here as a function of the effective anomaly $u$.
Full lines are the anisotropic NR predictions
(equation~\ref{eq:generic_grad}),
while the dashed ones are the pseudo-isotropic ones
(equation~\ref{def_PIso}).
Different colors correspond to different
background velocity anisotropies.
The test star's orbit is $\Jr = 0.1 \Lo$
and $L = 0.5 \Lo$
with $\Lo = \sqrt{G M b}$,
and it explores the cluster's core.
As in Fig.~\ref{fig:Ratio_NR_PIso},
for the present levels of anisotropies,
the differences between the two predictions
are minor.
}
\label{fig:local_vel_contrib}
\end{figure}

In this Appendix, we investigate the differences
in the anisotropic and pseudo-isotropic predictions
of the local velocity
deflections that accumulate as a test star
follows its unperturbed orbit.
Following equation~\eqref{eq:avgD_u},
this is best tracked by considering
$(\Delta v_{\parallel}^2)_{\mathrm{loc}}(u) = (\Delta v_{\parallel}^2)(r(u)) \, \Theta(u)$.
This is illustrated in Fig.~\ref{fig:local_vel_contrib}
for a test star orbiting within clusters' cores
with various background velocity anisotropies.
As already hinted in Fig.~\ref{fig:dFdt_Aniso_PIso},
for the range of anisotropies
considered here, we recover that the differences
between the fully anisotropic NR predictions
and the pseudo-isotropic ones
are only minor.

\balance


\begin{thebibliography}{}
\makeatletter
\relax
\def\mn@urlcharsother{\let\do\@makeother \do\$\do\&\do\#\do\^\do\_\do\%\do\~}
\def\mn@doi{\begingroup\mn@urlcharsother \@ifnextchar [ {\mn@doi@}
  {\mn@doi@[]}}
\def\mn@doi@[#1]#2{\def\@tempa{#1}\ifx\@tempa\@empty \href
  {http://dx.doi.org/#2} {doi:#2}\else \href {http://dx.doi.org/#2} {#1}\fi
  \endgroup}
\def\mn@eprint#1#2{\mn@eprint@#1:#2::\@nil}
\def\mn@eprint@arXiv#1{\href {http://arxiv.org/abs/#1} {{\tt arXiv:#1}}}
\def\mn@eprint@dblp#1{\href {http://dblp.uni-trier.de/rec/bibtex/#1.xml}
  {dblp:#1}}
\def\mn@eprint@#1:#2:#3:#4\@nil{\def\@tempa {#1}\def\@tempb {#2}\def\@tempc
  {#3}\ifx \@tempc \@empty \let \@tempc \@tempb \let \@tempb \@tempa \fi \ifx
  \@tempb \@empty \def\@tempb {arXiv}\fi \@ifundefined
  {mn@eprint@\@tempb}{\@tempb:\@tempc}{\expandafter \expandafter \csname
  mn@eprint@\@tempb\endcsname \expandafter{\@tempc}}}

\bibitem[\protect\citeauthoryear{{Bar-Or} \& {Alexander}}{{Bar-Or} \&
  {Alexander}}{2016}]{BarOr2016}
{Bar-Or} B.,  {Alexander} T.,  2016, \mn@doi [\apj]
  {10.3847/0004-637X/820/2/129}, \href
  {https://ui.adsabs.harvard.edu/abs/2016ApJ...820..129B} {820, 129}

\bibitem[\protect\citeauthoryear{{Binney} \& {Tremaine}}{{Binney} \&
  {Tremaine}}{2008}]{Binney2008}
{Binney} J.,  {Tremaine} S.,  2008, {Galactic Dynamics: Second Edition}.
Princeton Univ. Press

\bibitem[\protect\citeauthoryear{{Breen}, {Varri}  \& {Heggie}}{{Breen}
  et~al.}{2017}]{Breen+2017}
{Breen} P.~G.,  {Varri} A.~L.,   {Heggie} D.~C.,  2017, \mn@doi [\mnras]
  {10.1093/mnras/stx1750}, \href
  {https://ui.adsabs.harvard.edu/abs/2017MNRAS.471.2778B} {471, 2778}

\bibitem[\protect\citeauthoryear{{Brodie} \& {Strader}}{{Brodie} \&
  {Strader}}{2006}]{Brodie+2006}
{Brodie} J.~P.,  {Strader} J.,  2006, \mn@doi [\araa]
  {10.1146/annurev.astro.44.051905.092441}, \href
  {https://ui.adsabs.harvard.edu/abs/2006ARA&A..44..193B} {44, 193}

\bibitem[\protect\citeauthoryear{{Casertano} \& {Hut}}{{Casertano} \&
  {Hut}}{1985}]{Casertano+1985}
{Casertano} S.,  {Hut} P.,  1985, \mn@doi [\apj] {10.1086/163589}, \href
  {https://ui.adsabs.harvard.edu/abs/1985ApJ...298...80C} {298, 80}

\bibitem[\protect\citeauthoryear{{Chandrasekhar}}{{Chandrasekhar}}{1943}]{Chandrasekhar1943}
{Chandrasekhar} S.,  1943, \mn@doi [\apj] {10.1086/144517}, \href
  {https://ui.adsabs.harvard.edu/abs/1943ApJ....97..255C} {97, 255}

\bibitem[\protect\citeauthoryear{{Chavanis}}{{Chavanis}}{2012}]{Chavanis2012}
{Chavanis} P.-H.,  2012, \mn@doi [Physica A] {10.1016/j.physa.2012.02.019},
  \href {https://ui.adsabs.harvard.edu/abs/2012PhyA..391.3680C} {391, 3680}

\bibitem[\protect\citeauthoryear{{Chavanis}}{{Chavanis}}{2013a}]{Chavanis2013a}
{Chavanis} P.-H.,  2013a, \mn@doi [Eur. Phys. J. Plus]
  {10.1140/epjp/i2013-13126-9}, \href
  {https://ui.adsabs.harvard.edu/abs/2013EPJP..128..126C} {128, 126}

\bibitem[\protect\citeauthoryear{{Chavanis}}{{Chavanis}}{2013b}]{Chavanis2013}
{Chavanis} P.~H.,  2013b, \mn@doi [\aap] {10.1051/0004-6361/201220607}, \href
  {https://ui.adsabs.harvard.edu/abs/2013A&A...556A..93C} {556, A93}

\bibitem[\protect\citeauthoryear{{Cohn}}{{Cohn}}{1979}]{Cohn1979}
{Cohn} H.,  1979, \mn@doi [\apj] {10.1086/157587}, \href
  {https://ui.adsabs.harvard.edu/abs/1979ApJ...234.1036C} {234, 1036}

\bibitem[\protect\citeauthoryear{{Dejonghe}}{{Dejonghe}}{1987}]{Dejonghe1987}
{Dejonghe} H.,  1987, \mn@doi [\mnras] {10.1093/mnras/224.1.13}, \href
  {https://ui.adsabs.harvard.edu/abs/1987MNRAS.224...13D} {224, 13}

\bibitem[\protect\citeauthoryear{{Drukier}, {Cohn}, {Lugger}  \&
  {Yong}}{{Drukier} et~al.}{1999}]{Drukier+1999}
{Drukier} G.~A.,  {Cohn} H.~N.,  {Lugger} P.~M.,   {Yong} H.,  1999, \mn@doi
  [\apj] {10.1086/307243}, \href
  {https://ui.adsabs.harvard.edu/abs/1999ApJ...518..233D} {518, 233}

\bibitem[\protect\citeauthoryear{{Fouvry}, {Hamilton}, {Rozier}  \&
  {Pichon}}{{Fouvry} et~al.}{2021}]{Fouvry+2021}
{Fouvry} J.-B.,  {Hamilton} C.,  {Rozier} S.,   {Pichon} C.,  2021, \mn@doi
  [\mnras] {10.1093/mnras/stab2596}, \href
  {https://ui.adsabs.harvard.edu/abs/2021MNRAS.508.2210F} {508, 2210}

\bibitem[\protect\citeauthoryear{{Giersz} \& {Heggie}}{{Giersz} \&
  {Heggie}}{1994}]{Giersz+1994}
{Giersz} M.,  {Heggie} D.~C.,  1994, \mn@doi [\mnras]
  {10.1093/mnras/268.1.257}, \href
  {https://ui.adsabs.harvard.edu/abs/1994MNRAS.268..257G} {268, 257}

\bibitem[\protect\citeauthoryear{{Hamilton}, {Fouvry}, {Binney}  \&
  {Pichon}}{{Hamilton} et~al.}{2018}]{Hamilton+2018}
{Hamilton} C.,  {Fouvry} J.-B.,  {Binney} J.,   {Pichon} C.,  2018, \mn@doi
  [\mnras] {10.1093/mnras/sty2295}, \href
  {https://ui.adsabs.harvard.edu/abs/2018MNRAS.481.2041H} {481, 2041}

\bibitem[\protect\citeauthoryear{{Harris}}{{Harris}}{1991}]{Harris1991}
{Harris} W.~E.,  1991, \mn@doi [\araa] {10.1146/annurev.aa.29.090191.002551},
  \href {https://ui.adsabs.harvard.edu/abs/1991ARA&A..29..543H} {29, 543}

\bibitem[\protect\citeauthoryear{{Harris} \& {Racine}}{{Harris} \&
  {Racine}}{1979}]{Harris+1979}
{Harris} W.~E.,  {Racine} R.,  1979, \mn@doi [\araa]
  {10.1146/annurev.aa.17.090179.001325}, \href
  {https://ui.adsabs.harvard.edu/abs/1979ARA&A..17..241H} {17, 241}

\bibitem[\protect\citeauthoryear{{Heggie} \& {Hut}}{{Heggie} \&
  {Hut}}{2003}]{HeggieHut2003}
{Heggie} D.,  {Hut} P.,  2003, {The Gravitational Million-Body Problem}

\bibitem[\protect\citeauthoryear{{H{\'e}non}}{{H{\'e}non}}{1958}]{henon1958}
{H{\'e}non} M.,  1958, Annales d'Astrophysique, \href
  {https://ui.adsabs.harvard.edu/abs/1958AnAp...21..186H} {21, 186}

\bibitem[\protect\citeauthoryear{{H{\'e}non}}{{H{\'e}non}}{1961}]{Henon1961}
{H{\'e}non} M.,  1961, Annales d'Astrophysique, \href
  {https://ui.adsabs.harvard.edu/abs/1961AnAp...24..369H} {24, 369}

\bibitem[\protect\citeauthoryear{{H{\'e}non}}{{H{\'e}non}}{1971}]{Henon1971}
{H{\'e}non} M.~H.,  1971, \mn@doi [\apss] {10.1007/BF00649201}, \href
  {https://ui.adsabs.harvard.edu/abs/1971Ap&SS..14..151H} {14, 151}

\bibitem[\protect\citeauthoryear{{Heyvaerts}}{{Heyvaerts}}{2010}]{Heyvaerts2010}
{Heyvaerts} J.,  2010, \mn@doi [\mnras] {10.1111/j.1365-2966.2010.16899.x},
  \href {https://ui.adsabs.harvard.edu/abs/2010MNRAS.407..355H} {407, 355}

\bibitem[\protect\citeauthoryear{{Hong}, {Kim}, {Lee}  \& {Spurzem}}{{Hong}
  et~al.}{2013}]{Hong2013}
{Hong} J.,  {Kim} E.,  {Lee} H.~M.,   {Spurzem} R.,  2013, \mn@doi [\mnras]
  {10.1093/mnras/stt099}, \href
  {https://ui.adsabs.harvard.edu/abs/2013MNRAS.430.2960H} {430, 2960}

\bibitem[\protect\citeauthoryear{{Kim}, {Yoon}, {Lee}  \& {Spurzem}}{{Kim}
  et~al.}{2008}]{Kim2008}
{Kim} E.,  {Yoon} I.,  {Lee} H.~M.,   {Spurzem} R.,  2008, \mn@doi [\mnras]
  {10.1111/j.1365-2966.2007.12524.x}, \href
  {https://ui.adsabs.harvard.edu/abs/2008MNRAS.383....2K} {383, 2}

\bibitem[\protect\citeauthoryear{{Lightman} \& {Shapiro}}{{Lightman} \&
  {Shapiro}}{1978}]{Lightman+1978}
{Lightman} A.~P.,  {Shapiro} S.~L.,  1978, \mn@doi [Rev. Mod. Phys.]
  {10.1103/RevModPhys.50.437}, \href
  {https://ui.adsabs.harvard.edu/abs/1978RvMP...50..437L} {50, 437}

\bibitem[\protect\citeauthoryear{{Longaretti} \& {Lagoute}}{{Longaretti} \&
  {Lagoute}}{1997}]{Longaretti1997}
{Longaretti} P.~Y.,  {Lagoute} C.,  1997, \aap, \href
  {https://ui.adsabs.harvard.edu/abs/1997A&A...319..839L} {319, 839}

\bibitem[\protect\citeauthoryear{{Meylan} \& {Heggie}}{{Meylan} \&
  {Heggie}}{1997}]{Meylan+1997}
{Meylan} G.,  {Heggie} D.~C.,  1997, \mn@doi [\aapr] {10.1007/s001590050008},
  \href {https://ui.adsabs.harvard.edu/abs/1997A&ARv...8....1M} {8, 1}

\bibitem[\protect\citeauthoryear{{Rosenbluth}, {MacDonald}  \&
  {Judd}}{{Rosenbluth} et~al.}{1957}]{Rosenbluth+1957}
{Rosenbluth} M.~N.,  {MacDonald} W.~M.,   {Judd} D.~L.,  1957, \mn@doi [Phys.
  Rev.] {10.1103/PhysRev.107.1}, \href
  {https://ui.adsabs.harvard.edu/abs/1957PhRv..107....1R} {107, 1}

\bibitem[\protect\citeauthoryear{{Skowron} \& {Gould}}{{Skowron} \&
  {Gould}}{2012}]{Skowron+2012}
{Skowron} J.,  {Gould} A.,  2012, arXiv, \href
  {https://ui.adsabs.harvard.edu/abs/2012arXiv1203.1034S} {1203.1034}

\bibitem[\protect\citeauthoryear{{Spitzer}}{{Spitzer}}{1987}]{Spitzer1987}
{Spitzer} L.,  1987, {Dynamical evolution of globular clusters}.
Princeton Univ. Press

\bibitem[\protect\citeauthoryear{{Theuns}}{{Theuns}}{1996}]{Theuns1996}
{Theuns} T.,  1996, \mn@doi [\mnras] {10.1093/mnras/279.3.827}, \href
  {https://ui.adsabs.harvard.edu/abs/1996MNRAS.279..827T} {279, 827}

\bibitem[\protect\citeauthoryear{{Vasiliev}}{{Vasiliev}}{2015}]{Vasiliev2015}
{Vasiliev} E.,  2015, \mn@doi [\mnras] {10.1093/mnras/stu2360}, \href
  {https://ui.adsabs.harvard.edu/abs/2015MNRAS.446.3150V} {446, 3150}

\bibitem[\protect\citeauthoryear{{Wang}, {Spurzem}, {Aarseth}, {Nitadori},
  {Berczik}, {Kouwenhoven}  \& {Naab}}{{Wang} et~al.}{2015}]{Wang+2015}
{Wang} L.,  {Spurzem} R.,  {Aarseth} S.,  {Nitadori} K.,  {Berczik} P.,
  {Kouwenhoven} M.~B.~N.,   {Naab} T.,  2015, \mn@doi [\mnras]
  {10.1093/mnras/stv817}, \href
  {https://ui.adsabs.harvard.edu/abs/2015MNRAS.450.4070W} {450, 4070}

\makeatother
\end{thebibliography}
\end{document}